\documentclass[reprint,superscriptaddress,nobibnotes,amsmath,amssymb,aps,pra,floatfix]{revtex4-2}

\usepackage[utf8]{inputenc}
\usepackage{graphicx}
\usepackage{dcolumn}
\usepackage{bm}
\usepackage{upgreek}
\usepackage{physics}
\usepackage{hyperref}
\hypersetup{colorlinks = true, linkcolor = [rgb]{0.19411,0.51882,0.667058}, urlcolor = [rgb]{0.125490,0.29542,0.1647058}, citecolor = [rgb]{0.75882,0.37411,0.14117}}

\usepackage[caption=false]{subfig}
\usepackage{overpic}
\usepackage{xcolor}
\usepackage{sidecap}
\usepackage{amsmath}

\usepackage[T1]{fontenc}
\DeclareUnicodeCharacter{2215}{ }
\DeclareGraphicsExtensions{.pdf,.png}

\newcommand{\subfigim}[5][0.45\linewidth]{%
    \captionsetup[subfloat]{labelformat=empty}
    \subfloat[\label{#3}]{%
        \begin{overpic}[#1]{#2}
            \put(#4,#5){{(\thesubfigure)}}
        \end{overpic}
    }
}

\usepackage{fancyhdr}
\begin{document}

\title{Generating redundantly encoded resource states for photonic quantum computing}

\author{Samuel J. Sheldon}
\email{sam.sheldon@aegiq.com}
\affiliation{Aegiq Ltd., Cooper Buildings, Sheffield S1 2NS, United Kingdom}

\author{Pieter Kok}
\affiliation{Aegiq Ltd., Cooper Buildings, Sheffield S1 2NS, United Kingdom}
\affiliation{School of Mathematical and Physical Sciences, University of Sheffield, Sheffield S3 7RH, United Kingdom}

\date{\today}

\begin{abstract}
    \noindent
    Measurement-based quantum computing relies on the generation of large entangled cluster states that act as a universal resource on which logical circuits can be imprinted and executed through local measurements.
    A number of strategies for constructing sufficiently large photonic cluster states propose fusing many smaller resource states generated by a series of quantum emitters.
    However, the fusion process is inherently probabilistic with a 50\% success probability in standard guise.
    A recent proposal has shown that, in the limit of low loss, the probability of achieving successful fusion may be boosted to near unity by redundantly encoding the vertices of linear graph states using Greenberger-Horne-Zeilinger states \textbf{[Quantum 7, 992 (2023)]}.
    Here we present a protocol for deterministically generating redundantly encoded photonic resource states using single quantum emitters, and study the impact of protocol errors and photonic losses on the generated resource states and type-II photonic fusion.
    Our work provides a route for efficiently constructing complex entangled photonic qubit states for photonic quantum computing and quantum repeaters.
\end{abstract}

\maketitle

\section{Introduction}\label{sec:intro}\noindent
Photonic quantum computers are naturally suited to performing computations using the measurement-based model of quantum computing (MBQC)~\cite{raussendorf2001one, briegel2009measurement}.
In the MBQC model, first described by Raussendorf and Briegel \cite{raussendorf2001one}, a computation proceeds by first preparing a large entangled qubit state, known as a cluster state.
Local measurements are then used to imprint logical operations onto the cluster state, progress the computation through the entangled cluster state via teleportation, and read the outcome in the final measurement.
Recent attention has turned to using deterministic quantum emitters (QEs) to act as the generators of photonic cluster states.
However, the inherent properties of single QEs limit the size and dimensionality of the entangled photon states they can generate~\cite{pettersson2025deterministic}.
Hence entangled photonic qubit states with the requisite complexity of cluster states for universal~\cite{raussendorf2001one, nielsen2004optical, briegel2009measurement} and fault-tolerant~\cite{raussendorf2006fault} quantum computing, or even smaller states for photonic quantum repeaters~\cite{azuma2015all, buterakos2017deterministic}, cannot be generated by single quantum emitters alone.

\begin{figure}[t]
    \centering
    \includegraphics[width=7cm]{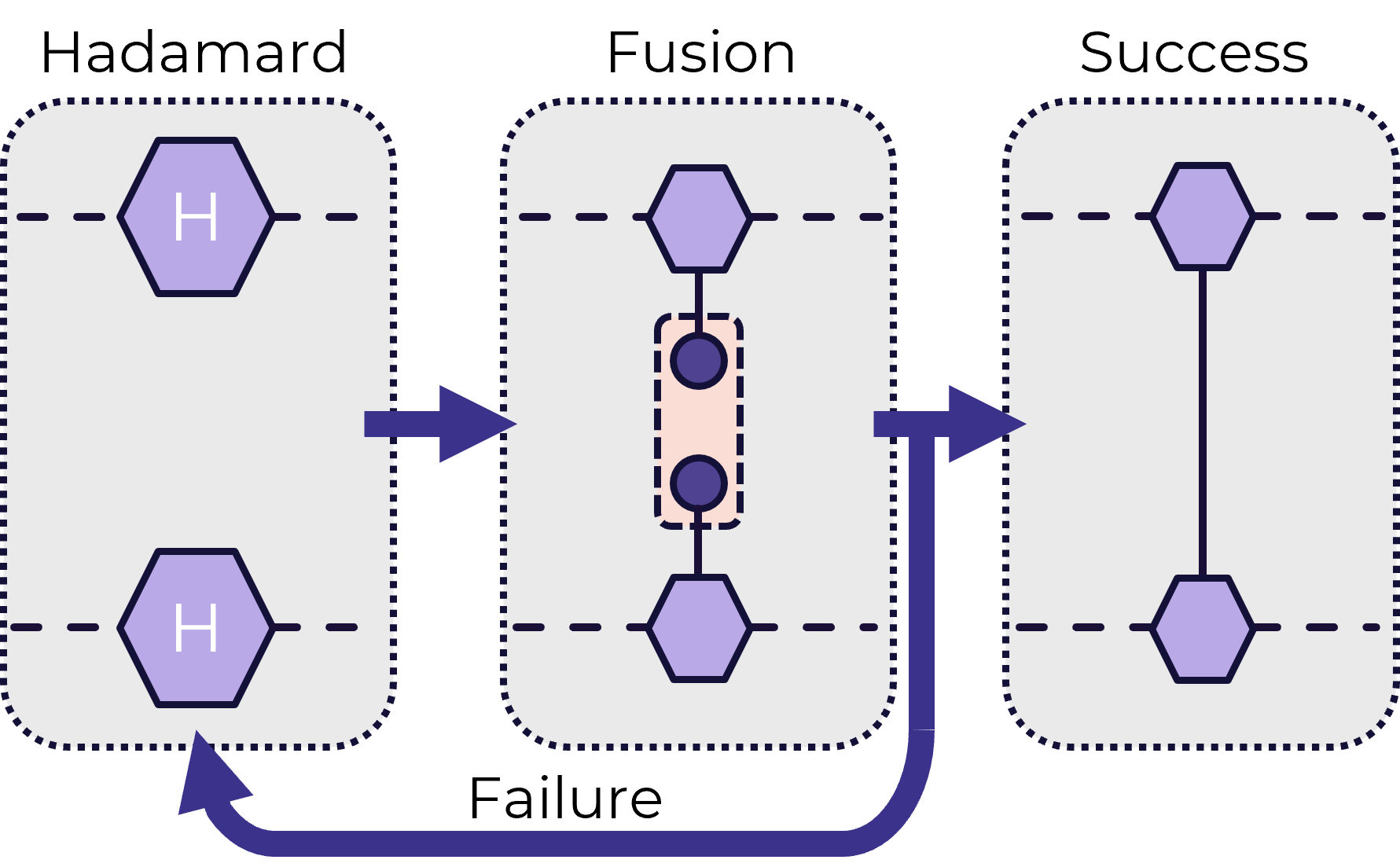}
    \caption{A diagram of the boosted fusion process proposed in~\cite{hilaire2023near}. Local Hadamard gates pull single photonic qubits (dark purple circles) out of the GHZ states (light purple hexagons) redundantly encoding logical qubits. The two photonic qubits pulled out by the Hadamard gates are consumed in a fusion measurement which, if the result indicates success, will generate bipartite entanglement between the logical qubits. If the fusion measurement fails the process is repeated.}
    \label{fig:boosted_fusion_process}
\end{figure}

To overcome this limitation, recent work has focused on generating large, multi-dimensional photonic cluster states by employing photonic fusion~\cite{browne2005resource, Bartolucci2023, wein2024minimizing, bartolucci2021creation} to join smaller resource states of fixed size that can in principle be generated by suitable single QEs even when accounting for limited emitter coherence times. 
However, the drawback of such an approach is the probabilistic and inefficient nature of this fusion process.
Photonic fusion in its standard guise generates entanglement with a 50\% probability, and consumes up to two photonic qubits per fusion operation regardless of the outcome~\cite{browne2005resource}.
A number of boosting schemes have been proposed to increase this success probability~\cite{lim2005repeat,grice2011arbitrarily, ewert20143, olivo2018ancilla, yamazaki2025linear, bartolucci2025comparison, hilaire2023near}, with one of particular interest here proposed by Hilaire et al.~\cite{hilaire2023near}.
This scheme combines redundantly encoded resource states, where each vertex of a linear graph state is replaced by a Greenberger-Horne-Zeilinger (GHZ) state, with a time-delayed repeat-until-success strategy to increase the fusion success probability.
Using this strategy photonic fusion is performed in two stages, as shown in Fig.~\ref{fig:boosted_fusion_process}.
First a local Hadamard gate is applied to a single redundantly encoded qubit in each of the pair of vertices involved in the fusion operation.
The Hadamard gate effectively draws these two redundantly encoded qubits out of their parent vertices to their own vertices in the graph representation of the state.
A fusion measurement may then be performed, consuming these qubits. The process is repeated until either the measurement outcome signifies successful fusion or no more redundantly encoding qubits remain.
In this way the fusion success probability can be boosted and the destruction of logical qubits carrying quantum information is avoided.

As vertices in the photonic resource states are redundantly encoded on GHZ states, this strategy is ultimately limited by photon loss, as any qubit loss entirely disrupts the entanglement within a GHZ state.
Optimising the fusion success probability thus requires balancing the degree of redundant encoding against optical loss.
Only when the end-to-end efficiency~\footnote{The probability of generating and successfully detecting a photon.} is greater than $\approx 82\%$ will increasing the degree of redundant encoding lead to increased fusion success probability~\cite{hilaire2023near}.

In this paper we propose a general protocol for generating redundantly encoded resource states of entangled time-bin photonic qubits, building on previous work employing QEs to generate either 1D graph states or GHZ states~\cite{denning2017protocol, lee2019quantum, vezvaee2022deterministic, hilaire2023near, tiurev2021fidelity, su2024continuous, Lindner2009}.
We show that implementing our protocol with a suitable QE results in the generation of a redundantly encoded resource state consisting of a series GHZ states linearly entangled in a 1D chain.
Furthermore, taking the example of two promising quantum dot (QD) systems, we examine how errors at each stage of our protocol impact the generated resource state, and derive expressions for the fidelities of the resulting states.

The paper is organised as follows.
In Section~\ref{sec:qd} we present a brief overview of two candidate QD systems we consider as potential generators of time-bin redundantly encoded photonic resource states.
We then outline our time-bin redundantly encoded resource state generation protocol in Section~\ref{sec:generation_protocol}.
Following this, in Section~\ref{sec:error}, we analyse the impact of error mechanisms at each step of our protocol and discuss the consequences for type-II photonic fusion.
We present our conclusions in Section~\ref{sec:conclusion}.

\section{Quantum dot systems}\label{sec:qd}\noindent
\begin{figure}
    \centering
    \hspace*{\fill}%
    \subfloat[\label{fig:single_qd_levels}]{%
         \includegraphics[height=4.5cm]{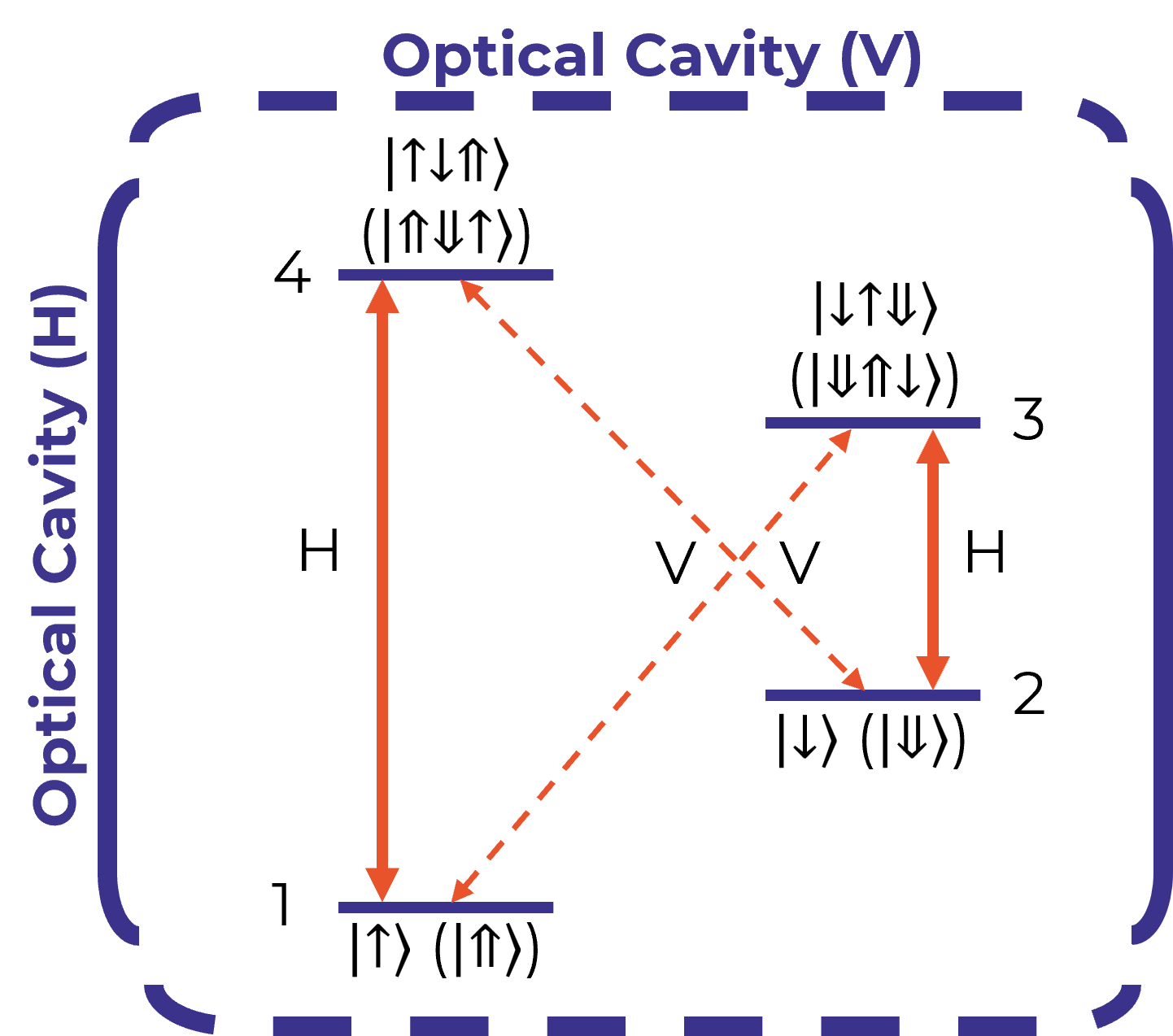}%
    }\hspace*{\fill}%
    \subfloat[\label{fig:qdm_levels}]{%
        \includegraphics[height=4.5cm]{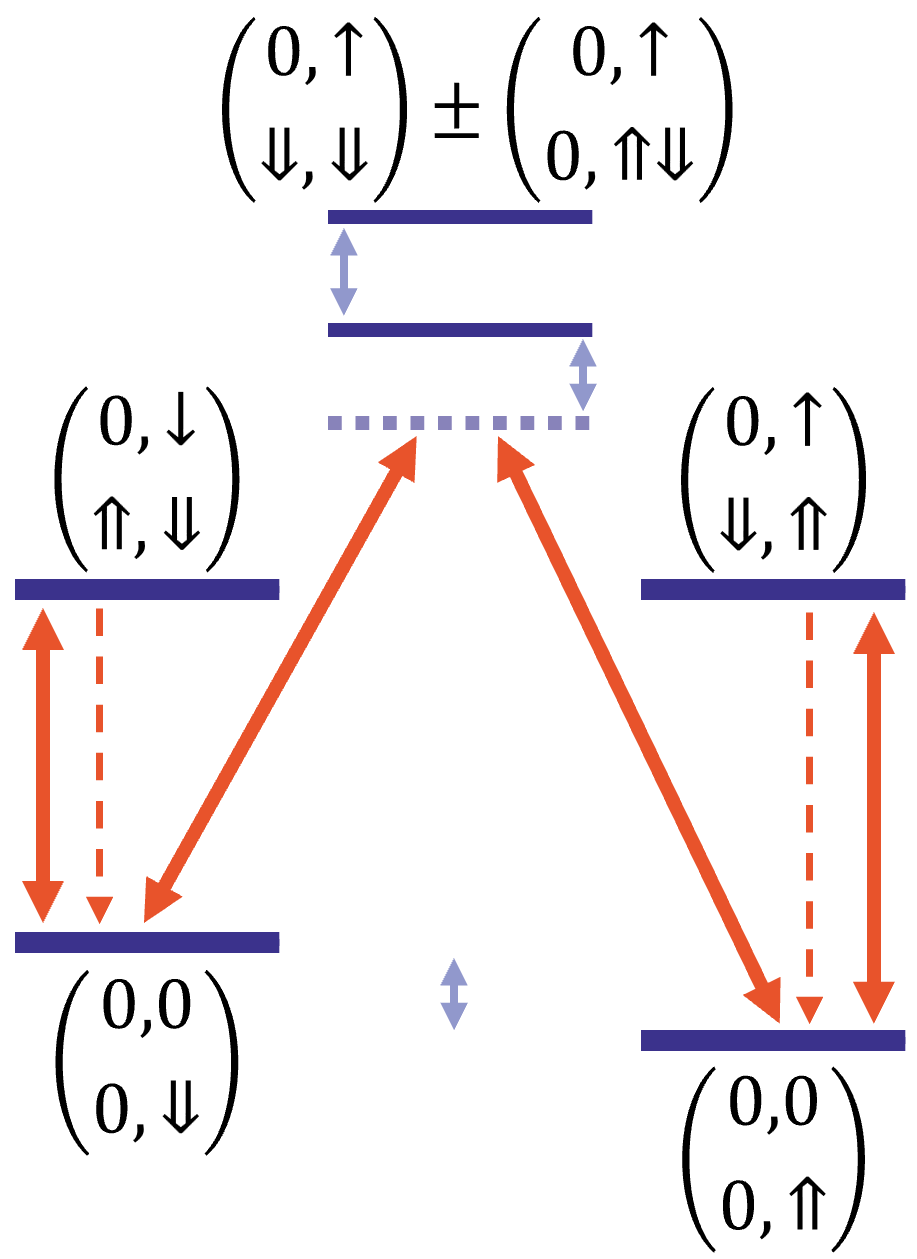}%
    }
    \hspace*{\fill}%
    \caption{The energy level structures of (a) a single charged QD situated in an in-plane magnetic field and (b) a charged QDM. Hole and electron states are denoted by $\{\Downarrow,\Uparrow\}$ and $\{\downarrow,\uparrow\}$ respectively. In (b) the notation $\begin{pmatrix} e_B & e_T \\ h_B & h_T\end{pmatrix}$ is used to denote the spatial position of electrons $e$ and holes $h$ in the top $T$ or bottom $B$ quantum dot of the QDM as in~\cite{vezvaee2022deterministic}.}
    \label{fig:qd_level_structure}
\end{figure}
Our protocol is applicable to any quantum emitter that possesses two ground states whose populations may be coherently transferred between one another and also possesses at least one individually-addressable cyclic transition~\footnote{A cyclic always returns the quantum emitter to the same ground state after excitation.} between one of the ground states and a single excited state.
Beyond determining the suitability of a quantum emitter to act as a resource state generator (RSG), the properties of the quantum emitter determine the nature of the errors that will inevitably arise in a physical implementation of our protocol (see Sec.~\ref{sec:error}).

We shall consider two quantum dot (QD) systems that have the necessary properties to generate redundantly encoded resource states, namely a single cavity-coupled charged QD situated in an in-plane magnetic field (Fig.~\ref{fig:single_qd_levels}), and a charged quantum dot molecule (QDM) (Fig.~\ref{fig:qdm_levels}).
Both systems may be deterministically occupied by a single charge carrier (electron or hole) leading to $\Lambda$-type energy level structures (see Fig.~\ref{fig:qd_level_structure}), arising from the in-plane magnetic field in the single QD system~\cite{Warburton2013} and hole spin mixing in the QDM~\cite{doty2010hole, economou2012scalable}.
This enables the populations of the spin-qubit states (the two single charge-carrier states) to be coherently transferred via a Raman transition induced by detuned optical excitation~\cite{Warburton2013, Majumdar2013, vezvaee2023avoiding}.
Combining the Raman transition with a natural precession of the spin-qubit state about the $z$-axis of the Bloch sphere (resulting from an energy splitting between its two basis states) gives access to all states on the Bloch sphere.
Here the $z$-axis is defined along the magnetic field axis in the single QD system and along the optic axis in the QDM and the spin states are written in the basis along this axis.

In both QD systems photon generation occurs via spontaneous emission.
Individual transitions can be addressed optically using carefully designed optical excitation pulses.
Optical excitation causes a transfer of population between the ground and excited states coupled by the driven transition.
Decay of the excited state back to the ground state(s) is accompanied by spontaneous emission of a photon.
Careful control of the ground state populations enables these QEs to generate photonic qubits encoded in the time-bin basis (see Sec.~\ref{sec:generation_protocol}), whereby time is segmented into early and late bins that map to the computational basis $\{\ket{0},\ket{1}\}$.
Using cyclical transitions for photon generation ensures that after excitation the system returns to its initial ground state avoiding unwanted spin flips.
QDMs naturally possess the cyclic transitions required for photon generation~\cite{vezvaee2022deterministic}.
In the absence of an applied magnetic field, or when a field is applied out-of-plane, charged single QDs also possess cyclic transitions.
However, the application of the in-plane magnetic field typically used to enable coherent optical control of the spin state removes these cyclic transitions.
Quasi-cyclic transitions may be recovered in this case via coupling to an optical structure~\cite{PhysRevLett.113.093603, reinhard2012strongly, ohta2011strong} to induce a relative enhancement of the decay rate of one (pair of) optical transition(s)~\cite{sheldon2024optical}. 
We note that recent work has demonstrated high-fidelity control of electron spins combined with highly cyclical optical transitions in single QDs situated in an out-of-plane magnetic field by using light-hole mixing to recover a similar energy level structure to the in-plane magnetic field case~\cite{koong2025coherent}.

The upper limit on the size of the resource states that can be reliably generated by a single quantum emitter is ultimately determined by its coherence time and the error mechanisms to which it and the generated photonic state are subject.
A number of experiments have successfully demonstrated generation of entangled photonic qubits.
Using quantum dot platforms, generation of linear graph states with up to 10 polarisation-encoded photonic qubits has been experimentally demonstrated~\cite{cogan2023deterministic}, as has reconfigurable generation of 4-qubit spin-photon entangled qubit states with redundant encoding~\cite{huet2025deterministic} and 4-qubit photonic GHZ states with fidelities of approximately 84\%~\cite{pont2024high}.

\section{Time-bin redundantly encoded resource state generation protocol}\label{sec:generation_protocol}\noindent
\begin{figure*}[ht!]
    \centering
    \hspace*{\fill}
    \subfloat[Step 1(a)\label{fig:qd_initialisation}]{%
         \includegraphics[height=2.5cm]{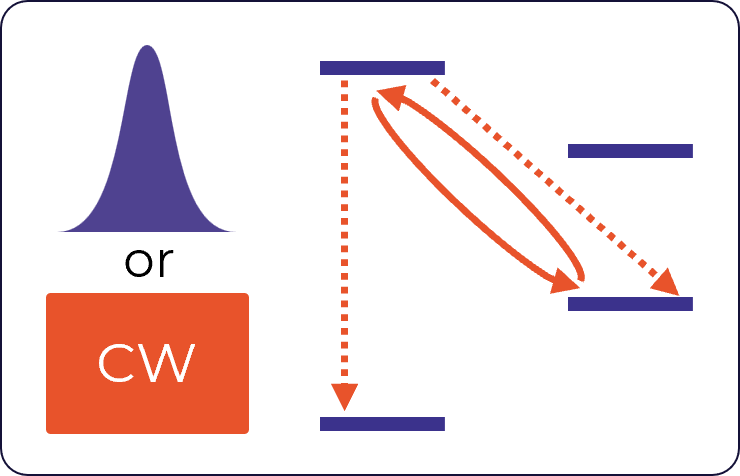}%
    }\hspace*{\fill}%
    \subfloat[Step 1(b)\label{fig:qd_control_state_prep}]{%
         \includegraphics[height=2.5cm]{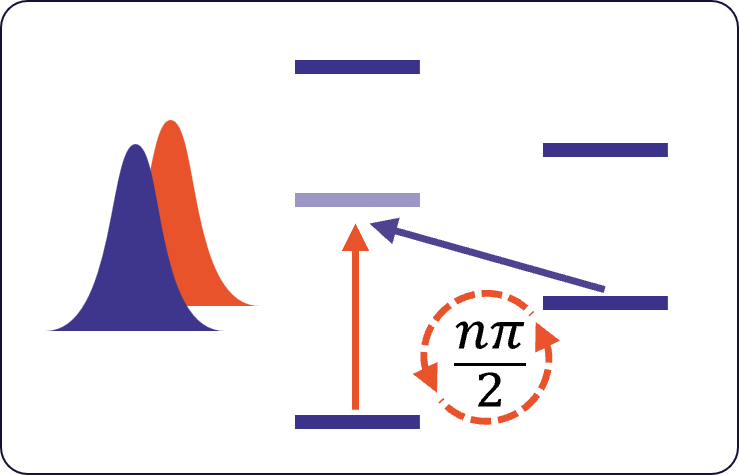}%
    }\hspace*{\fill}%
    \subfloat[Step 2\label{fig:qd_excitation_early}]{%
         \includegraphics[height=2.675cm]{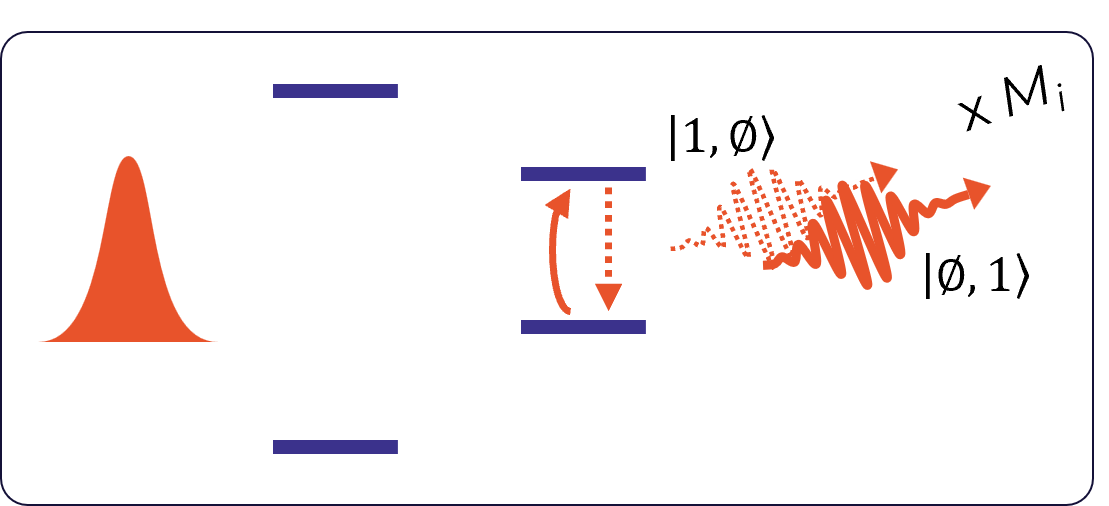}%
    }\hspace*{\fill}

    \hspace*{\fill}
    \subfloat[Step 3\label{fig:qd_state_flip}]{%
         \includegraphics[height=2.5cm]{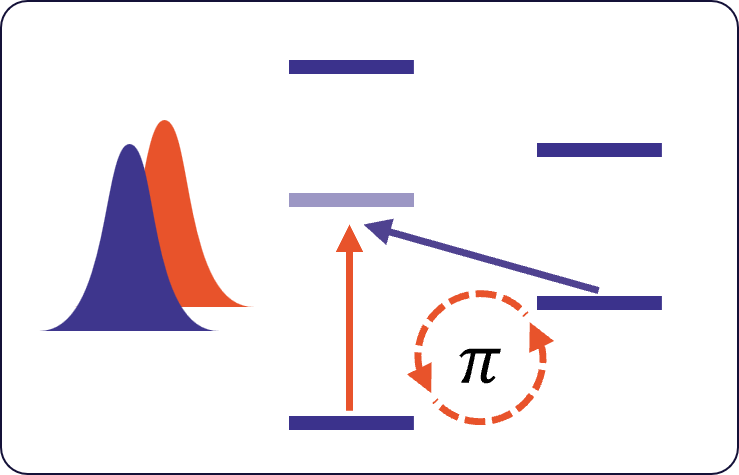}%
    }\hspace*{\fill}%
    \subfloat[Step 4\label{fig:qd_excitation_late}]{%
         \includegraphics[height=2.675cm]{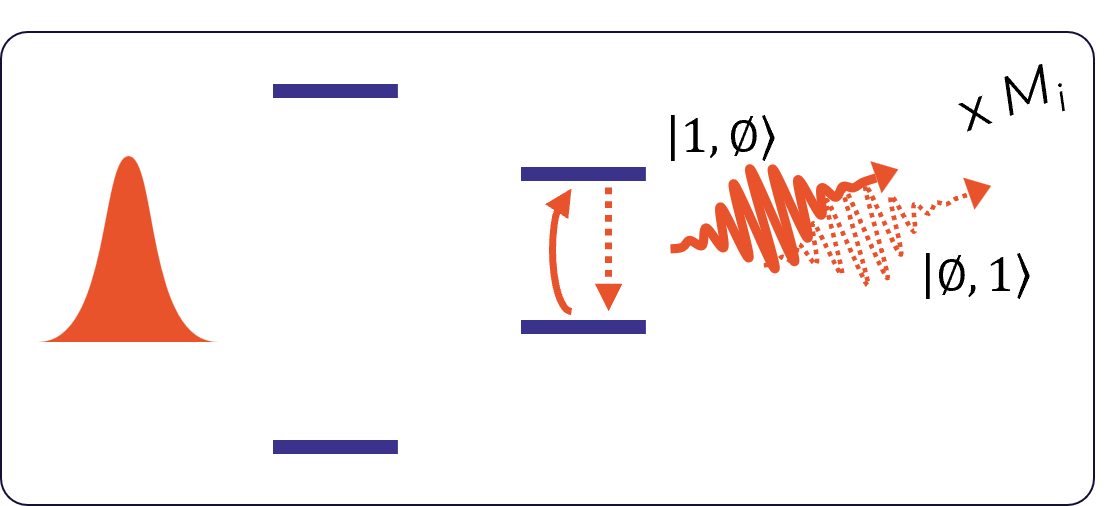}%
    }\hspace*{\fill}%
    \subfloat[Step 5\label{fig:qd_control_entanglement}]{%
         \includegraphics[height=2.5cm]{diagrams/protocol_control_process_spin_hadamard}%
    }\hspace*{\fill}
    \caption{Depictions of the steps of the resource state generation protocol using the example of a charged single QD as the resource state generator. (a) Step 1a prepares the resource state generator (RSG) in a known state via known initialisation protocols~\cite{sheldon2024optical}. (b) In Step 1b control pulses prepare the RSG in a 50:50 superposition of the two ground states ($n\in\{1,3\}$). (c) In step 2 $M_i$ sequential excitation pulses incident on the RSG cause $M_i$ photons to be generated if the RSG initially occupies the ground state selected to be the bright state. (d) Step 3 inverts the populations of the ground states. (e) In step 4 a second set of $M_i$ sequential optical pulses cause $M_i$ photons to be sequentially generated if the RSG occupies the ground state selected to be the bright state after the prior control pulse. (f) A further control pulse prepares the RSG to generate the next photonic qubit(s) and determines the nature of the entanglement between photonic qubits in step 5 ($n\in\{1,2,3\}$).}
    \label{fig:protocol_steps}
\end{figure*}
We now outline our protocol for generating time-bin redundantly encoded photonic resource states building on prior protocols for GHZ and 1D graph state generation~\cite{denning2017protocol, lee2019quantum, vezvaee2022deterministic, hilaire2023near, tiurev2021fidelity}. 
The steps of our protocol can be broadly separated into three categories.
Preparation of the RSG for generating entangled photonic qubit states occurs in step 1.
Steps 2-4 are responsible for photonic qubit generation, either as an individual qubit or multiple qubits in a GHZ state.
Step 5 determines the nature of the entanglement between the photonic qubit(s) generated in the current round of the protocol and the photonic qubit(s) generated in the next round of the protocol.
Using the notation $\ket{\rm late,~early}$ to indicate the occupancy of early and late time-bins, our protocol is as follows:
\begin{enumerate}
    \item Prepare the spin-qubit in a 50:50 superposition~\footnote{Or a non-50:50 superposition to generate the first vertex in the same non-50:50 superposition state.} of its basis states by:
    \begin{enumerate}
        \item initialising the spin-qubit in one of its basis states (Fig.~\ref{fig:qd_initialisation}), and
        \item applying either a spin-Hadamard $R_y(\pi/2)$ or inverse spin-Hadamard $R_y(3\pi/2)$ gate where $R_y(\theta)=\exp{-iY\theta/2}$ is the $y$-rotation operator (Fig.~\ref{fig:qd_control_state_prep}),
    \end{enumerate}
    \item Drive the selected cycling transition with $M_i$ consecutive pulses to generate $M_i$ photons in early time-bins $\bigotimes_{M_i}\ket{\oslash,1}$ (Fig.~\ref{fig:qd_excitation_early}). Photon generation is conditional on the occupation of the ground state of the driven transition,
    \item Invert the populations of the spin-qubit basis states by application of a spin-flip $R_y(\pi)$ (Fig.~\ref{fig:qd_state_flip}),
    \item Drive the selected cycling transition with $M_i$ consecutive pulses to generate $M_i$ photons in late time-bins $\bigotimes_{M_i}\ket{1,\oslash}$. Photon generation is again conditional on the occupation of the ground state of the driven transition (Fig.~\ref{fig:qd_excitation_late}),
    \item Either (Fig.~\ref{fig:qd_control_entanglement}):
    \begin{enumerate}
        \item invert the populations of the spin states and repeat steps (2)-(4) to further redundantly encode the vertex on additional time-bin photonic qubits ($i>1$), or
        \item to complete encoding of a vertex of the state on $M_n=\sum_iM_i$ photonic qubits, for all subsequent cycles of the protocol either:
        \begin{enumerate}
            \item consistently apply a spin-Hadamard gate or inverse spin-Hadamard gate when, in step (1b), the spin was prepared in the $\ket{+}$ or $\ket{-}$ state respectively, or
            \item alternate between applying inverse spin-Hadamard and spin-Hadamard gates if the spin was prepared in the $\ket{+}$ state in step (1b), reversing the order if prepared in the $\ket{-}$ state instead,
        \end{enumerate}
    \end{enumerate}
    \item Repeat steps (2)-(5) $N$ times.
\end{enumerate}

Taking the example where a single photon can be generated only when the spin-qubit is in the spin-down state and using step (5bi) to complete redundant encoding of the photonic vertices, the resulting entangled spin-photon state (neglecting a global phase) is

\begin{equation}
    \begin{split}
    \ket{\Psi_{\rm RE}}_N=&~\mathbf{CZ}_{\uparrow,\mathcal{V}_N}\Bigg(\prod_{n=1}^{N-1}\mathbf{CZ}_{\mathcal{V}_{n+1},\mathcal{V}_{n}}\Bigg)\\
    &\times\frac{1}{\sqrt{2}}\Big(\ket{\downarrow}\pm\ket{\uparrow}\Big)
    \bigotimes_{n=1}^N\ket{-}_{n,M_n},
    \end{split}
    \label{eq:re_spin_photon_state_single_control}
\end{equation}
where the photonic vertices are prepared in the state
\begin{align}\nonumber
    \ket{\pm}_{n,M_n} = \frac{1}{\sqrt{2}}\Bigg(\bigotimes_{m=1}^{M_n}\ket{\oslash,1}_{n,m}\pm\bigotimes_{m=1}^{M_n}\ket{1,\oslash}_{n,m}\Bigg)     
\end{align}
with the subscripts $n$ and $m$ indicating the $m^{\rm th}$ photonic qubit in the $n^{\rm th}$ vertex of the state.
If step (5bii) is instead used to complete the redundant encoding process then the resulting state is
\begin{equation}
\begin{split}
    \ket{\Psi_{\rm RE}}_{N} =&~\mathbf{CZ}_{\uparrow,\mathcal{V}_N}\Bigg(\prod_{n=1}^{N-1} \mathbf{CZ}_{\mathcal{V}_{n+1},\mathcal{V}_{n}}\Bigg)\\
    &\times\frac{1}{\sqrt{2}}\Big(\ket{\downarrow} \pm (-1)^{N}\ket{\uparrow}\Big)
    \bigotimes_{n=1}^N\ket{+}_{n,M_n}.
\end{split}
\label{eq:re_spin_photon_state_alternating_control}
\end{equation}
The CZ operators account for bipartite entanglement between neighbouring vertices indicated by the subscripts, imparting a phase of $-1$ to the $\ket{\uparrow}\otimes_{M_N}\ket{1,\oslash}$ and $\otimes_{M_{n+1}}\ket{1,\oslash}\otimes_{M_n}\ket{1,\oslash}$ components of the state.
The $\pm$ in the spin-qubit component of the state corresponds to the initial spin-qubit superposition state created at step (1b) of the protocol.
Eqs.~\eqref{eq:re_spin_photon_state_single_control} and~\eqref{eq:re_spin_photon_state_alternating_control} thus describe a state where a spin-qubit is entangled with the final vertex of a redundantly encoded photonic resource state $\mathcal{R}$  consisting of $N$ vertices $\mathcal{V}_n\in\mathcal{R}$ each entangled with their nearest neighbours and redundantly encoded on $M_n$ time-bin photonic qubits entangled in a GHZ state (in Appendix~\ref{app:dual_rail_encoded_resource_state} we present a modification to the protocol for generating resource states of polarisation / dual-rail encoded photonic qubits).
With an appropriate choice of encoding, we may rewrite Eqs.~\eqref{eq:re_spin_photon_state_single_control} and~\eqref{eq:re_spin_photon_state_alternating_control} in the computational basis as
\begin{equation}
\begin{split}
    \ket{\Psi_{\rm RE}}_N\equiv &
    \mathbf{CZ}_{\uparrow,N}\Bigg(\prod_{n=1}^{N-1}\mathbf{CZ}_{n+1,n}\Bigg)\ket{+}_S\\
    &\bigotimes_{n=1}^{N}\frac{1}{\sqrt{2}}\Bigg(\bigotimes_{m=1}^{M_n}\ket{0}_{n,m} + \bigotimes_{m=1}^{M_n}\ket{1}_{n,m}\Bigg)
\end{split}
\label{eq:state_vector_logical_basis}
\end{equation}
where $\ket{+}_S=(\ket{0}_S+\ket{1}_S)/\sqrt{2}$ accounts for the RSG.
This state is a generalisation of both the GHZ and 1D linear cluster states~\cite{hilaire2023near, pettersson2025deterministic}.
In the limit of $N=1$, $M_1>1$ $\ket{\Psi}$ is an $M_1+1$ qubit GHZ state, while when $M_n=1~\forall~n$, $N>0$ $\ket{\Psi}$ is an $N+1$ 1D linear graph state.
Note that while throughout this paper we shall assume photons are generated when driving the $H$-polarised transition coupling the spin-down state to the relevant trion state, the protocol will generate similar states if a different transition and/or spin state is selected for photon generation.

Applying multiple consecutive excitation pulses at steps (2) and (4) generates sub-vertices $\mathcal{S}_{n,k}\in\mathcal{V}_n$ within the redundantly encoded vertex $\mathcal{V}_n$ of the state (see Fig.~\ref{fig:resource_state_ghz}).
In the absence of errors multiple instances of these sub-vertices are entangled such that they form a single larger GHZ state.
While this may initially make the distinction between vertices and sub-vertices seem insignificant, the ordering of early and late time-bins within the sub-vertices is modified by this excitation pulse sequence.
Rather than the late time-bin of the $m^{\rm th}$ photonic qubit immediately following its early time-bin counterpart, the time-bins of all photonic qubits within the sub-vertex are grouped such that all of the early time-bins are first generated followed by all of the late time-bins (see Fig.~\ref{fig:vertex_time_bin_ordering}).
This unconventional temporal ordering of time-bins may prove beneficial when converting from time-bin to dual-rail encoding as required to perform local operations, such as Hadamard gates, on individual photonic qubits.
Generating the time-bins of a group of photonic qubits in blocks reduces the number of routing operations required in the conversion between time-bin and dual-rail encoding potentially allowing slower hardware with lower optical losses to be used.

Coherent control of the phase associated with each spin-qubit basis state, and the manner in which this is transferred to the photonic qubits, underpins the generation of the entangled spin-photon states.
During steps (2) to (4) of the protocol the spin of the QE becomes entangled with the newly generated photonic qubits in a single spin-photon GHZ state.
Step (5b) then applies a local operation that effectively pulls the spin-qubit out of this GHZ state to its own vertex in the state ready for additional photonic qubit generation.
We choose the control operations and their ordering to ensure the relative phase between the early and late components of the redundantly encoded photonic vertices remains constant throughout the state.
Deviating from the order of operations outlined in our protocol can lead to this phase becoming dependent on the position of the qubit within the state which would need to be accounted for via select application of phase gates or varying logical encoding.
A consequence of our control pulse sequence is that the internal phase of the spin-qubit can vary depending on the size of the state.
However, once disconnected from the photonic component of the state (via direct measurement or measurement of the final photonic vertex) this phase becomes inconsequential.
We note that after the spin-qubit is disconnected from the photonic component of the state a remnant phase originating from the severed bipartite entanglement can remain associated with the late component of the final vertex of the resource state.
This can be accounted for by the local application of a phase gate to this component of the state to return the redundantly encoded resource state to canonical graph form at the logical level.

\begin{figure}
    \centering
    \vspace*{-0.5cm}
    \subfigim[width=7cm]{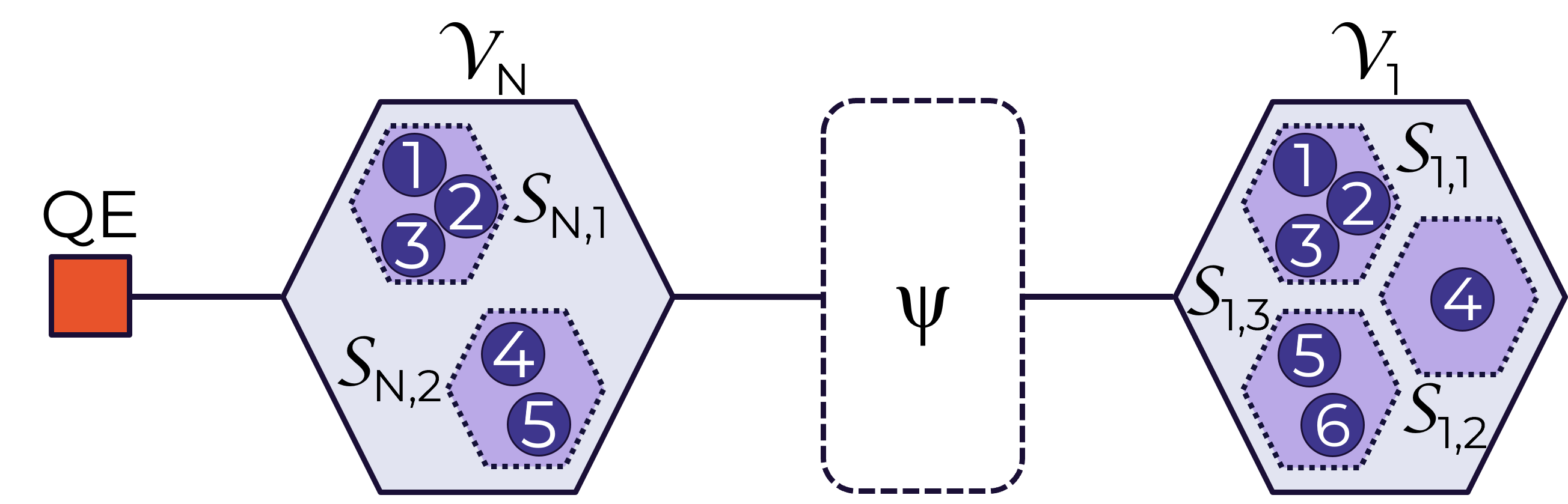}{fig:resource_state_ghz}{0}{28}
    \vspace*{-0.5cm}
    \subfigim[width=7cm]{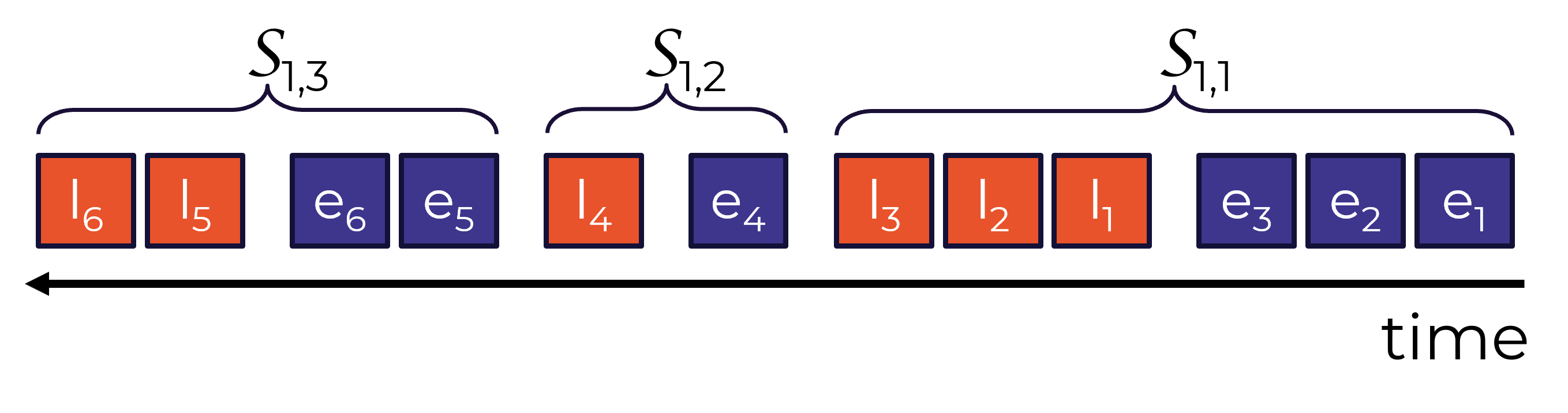}{fig:vertex_time_bin_ordering}{0}{25}
    
    \caption{
    (a) A representative example of a redundantly encoded resource state entangled with a quantum emitter (QE). Each vertex of the photonic component of the state is redundantly encoded on a GHZ state (large hexagon) and entangled with its nearest neighbours. Each of the GHZ states is formed by a number of sub-vertices (small hexagons) which in turn are formed by individual photonic qubits (circles) entangled in a GHZ state (where appropriate). (b) The temporal ordering of the time-bins of the photonic qubits in $\mathcal{V}_1$ in (a).}
    \label{fig:resource_state_example}
\end{figure}

\section{Impact of error mechanisms on resource state generation}\label{sec:error}\noindent
So far in our discussions we have assumed ideal circumstances.
However, any physical implementation of a RSG will naturally encounter errors, the nature of which will depend on the quantum system used as the RSG.
Taking the examples of the physical RSG systems outlined in Sec.~\ref{sec:qd}, when a QDM acts as the RSG the system can be subject to imperfect spin control, inefficient excitation, and photon loss.
If instead a cavity-coupled single QD situated in an in-plane magnetic field acts as the RSG additional error mechanisms beyond those already listed, namely imperfect cyclic transitions and off-resonant excitation of undesired optical transitions, also become relevant.
In this section, we individually study the impact of these error mechanisms on the generated spin-photon entangled state using the implementation of the protocol designed to generate the photonic vertices in the $\ket{+}$ state.
We also discuss the consequences of these errors on subsequent type-II fusion operations after the individual qubits are converted from time-bin to dual-rail encoding (see Appendix~\ref{app:fusion} for a brief review of type-II photonic fusion).

To quantify the impact of the studied error mechanisms on the generated entangled spin-photon state we use the fidelity to the ideal state in Eq.~\eqref{eq:re_spin_photon_state_alternating_control}.
The fidelity of the generated state $\ket{\Phi}$ and target states $\ket{\Psi}$ is given by
\begin{equation}
    \mathcal{F} = \bra{\Psi}\sigma\ket{\Psi}
    \label{eq:fidelity_def}
\end{equation}
where $\sigma=\ketbra{\Phi}{\Phi}$ is the density matrix of the generated state.

\subsection{QD state preparation errors}

Perhaps the least impactful of the possible error mechanisms is improper preparation of the RSG spin-qubit state in step (1) of the protocol.
We consider the scenario where, rather than preparing the spin-qubit in one of its basis states, step (1a) prepares the spin-qubit in the mixed state $(\abs{\alpha}^2\ketbra{\downarrow}{\downarrow} + \abs{\beta}^2\ketbra{\uparrow}{\uparrow})$ with $\abs{\alpha}^2 + \abs{\beta}^2=1$ as can occur with initialisation via population shelving~\cite{sheldon2024optical}, and the control operation in step (1b) experiences both $y-$ and $z-$rotation errors implementing the rotation $R_z(\Delta_z)R_y(n\pi/2+\Delta_y)$ where $n=1~(n=3)$ for a (inverse) spin-Hadamard gate.
The density matrix describing the resulting state is given by
\begin{equation}
    \sigma_{\rm RE} = \mathcal{F}_{s}\ketbra{\Phi_{\rm RE}}{\Phi_{\rm RE}} + (1-\mathcal{F}_{s})\ketbra{\Phi'_{\rm RE}}{\Phi'_{\rm RE}}.
\end{equation}
Here $\mathcal{F}_{s} = \bra{s}\frac{1}{2}(\abs{\alpha}^2\ketbra{\downarrow}{\downarrow} + \abs{\beta}^2\ketbra{\uparrow}{\uparrow})\ket{s}$  is the fidelity of the prepared spin state to the target spin state $s\in\{\uparrow,\downarrow\}$, $\ket{\Phi_{\rm RE}}$ is the resulting state when the spin-qubit was correctly prepared, and $\ket{\Phi'_{\rm RE}}$ is the resulting state when spin-qubit was prepared in the orthogonal state (see Appendix~\ref{app:initialisation_error}). 

\begin{figure}[t]
    \centering
    \vspace*{-0.5cm}
    \hspace*{\fill}
    \subfloat[\label{fig:init_rotation_error_step_1}]{%
         \includegraphics[width=4.4cm]{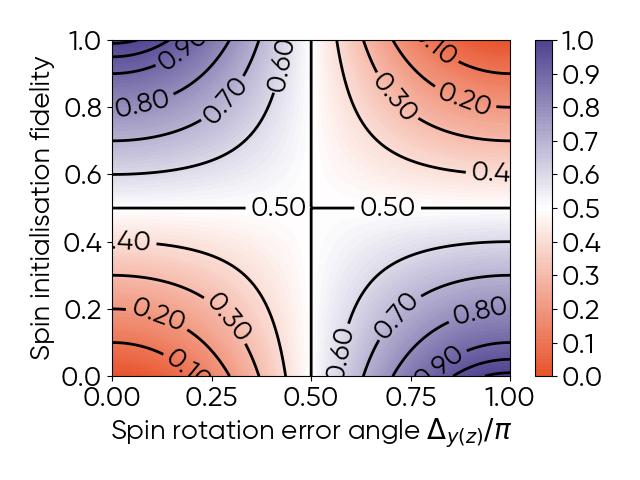}%
    }\hspace*{\fill}%
    \subfloat[\label{fig:rotation_error_step_1}]{%
         \includegraphics[width=4.4cm]{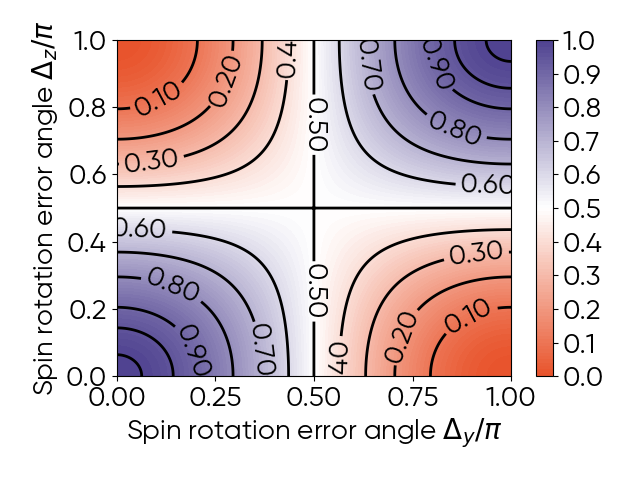}%
    }\hspace*{\fill}%
    
    \caption{The calculated fidelity of the spin-photon resource state when the spin preparation process in step (1) of the protocol is subject to (a) spin initialisation and either a $y$-rotation or $z-$rotation error, and (b) step (1a) is performed with unity fidelity but step (1b) is subject to both a $y$-rotation and $z$-rotation error.}
    \label{fig:spin_prep_error_state_fidelity}
\end{figure}

Using Eq.~\eqref{eq:fidelity_def} the fidelity of the resulting state to the ideal state is found to be
\begin{equation}
    \mathcal{F} = \frac{1}{2}\Big\{\Big(2\mathcal{F}_{s} - 1\Big)\cos{\Delta_y}\cos{\Delta_z} + 1\Big\}.
\end{equation}
We plot this spin-photon state fidelity, setting the fidelity of one of the three error mechanisms to unity, in Fig.~\ref{fig:spin_prep_error_state_fidelity}.

Beyond the calculated fidelity, the resulting spin-photon state reveals that errors resulting from spin-qubit preparation errors may be readily accounted for.
Removal of the first vertex from the photonic state, either via direct measurement or measurement of the second vertex in the computational basis, will improve the fidelity of the remaining state.
Alternatively, performing spin initialisation via the generation and measurement of a single photonic qubit entangled with the spin qubit in a Bell state (by following steps 1-4 of the protocol) will project the spin qubit into a pure state negating the first of the two spin preparation errors.
The measurement outcome will determine either the control operation applied in step (1b) or if an additional spin-flip operation is required to prepare the spin-qubit in the target state.

\subsection{Spin control errors}

Optical control of the RSG state in steps (3) and (5) of the protocol may similarly be subject to error in a physical system.
The effect of these imperfections is dependent on the control process and the step of the protocol impacted by the imperfection.

\subsubsection{Photonic qubit generation error}

\begin{figure}[t!]
    \centering
    \vspace*{-0.75cm}
    \subfigim[width=4.5cm]{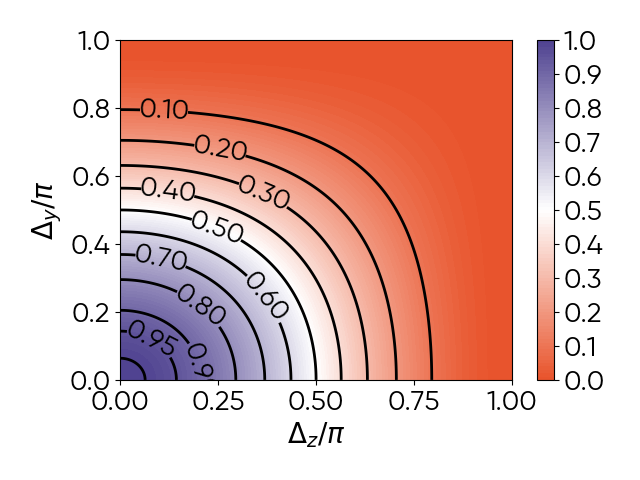}{fig:flip_error_a}{68}{60}
    \hspace*{-.3cm}
    \subfigim[width=4.5cm]{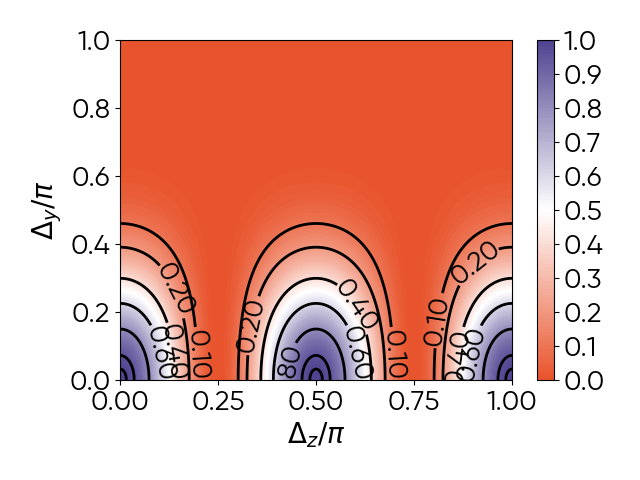}{fig:flip_error_b}{68}{60}
    \vspace*{-1cm}
    \subfigim[width=4.5cm]{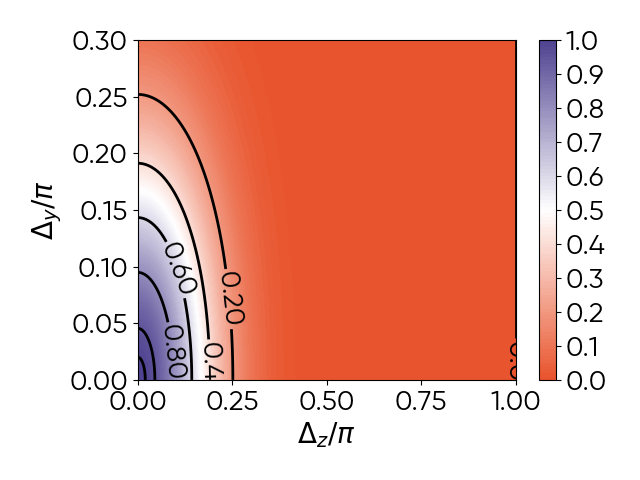}{fig:flip_error_c}{68}{60}
    \hspace*{-.25cm}
    \subfigim[width=4.5cm]{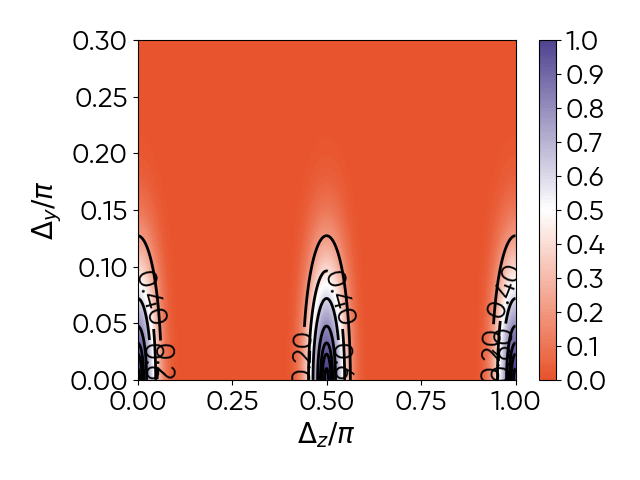}{fig:flip_error_d}{68}{60}
    \caption{The calculated spin-photon state fidelity when accounting for spin flip errors in steps (3) and (5a) of the protocol assuming each operation experiences the same $y$- and $z$-rotation error (i.e. $\Delta_{y(z)}^{(n,j)} = \Delta_{y(z)}~\forall~n,j$) for (a) 1 vertex encoded by 1 sub-vertex, (b) 1 vertex encoded by 4 sub-vertices, (c) 10 vertices each encoded by 1 sub-vertex, and (d) 10 vertices each encoded by 4 sub-vertices.}
    \label{fig:spin_flip_error}
\end{figure}

We begin by considering the impact of spin flip errors in step (3) of the protocol.
This spin flip operation ensures one and only one photon is generated per photonic qubit in a particular superposition of the early and late time-bins determined by the superposition state of the spin-qubit.
Consequently, an error in step (3) of the form $R_z(\Delta_z)R_y(\pi+\Delta_y)$ alters the composition of the photonic qubits in the state.
This error introduces a non-zero probability of a given (sub-) vertex prepared in a superposition of a vacuum state and two-photon state $\ket{\oslash,\oslash} \pm\ket{1,1}$ and reduces the probability of producing the given (sub-) vertex in the desired single-qubit or GHZ state as shown in Appendix~\ref{app:step_3_error}.
The fidelity of the resulting state to the ideal state is given by
\begin{equation}
    \mathcal{F} = \abs{\prod_n\Bigg\{\cos\Bigg(\sum_j\frac{\Delta_z^{(n,j)}}{2}\Bigg)\prod_j\cos\left({\frac{\Delta_y^{(n,j)}}{2}}\right)\Bigg\}}^2
    \label{eq:fidelity_spin_flip}
\end{equation}
where $\Delta_{y(z)}^{(n,j)}$ is the $y$($z$) error on the spin flip operation when generating the $j^{\rm th}$ sub-vertex redundantly encoding the $n^{\rm th}$ vertex.
This fidelity is plotted in Fig.~\ref{fig:spin_flip_error} assuming each spin flip operation experiences the same $y-$ and $z-$error.
When a single vertex is encoded on more than one sub-vertex our results show peaks in the fidelity when $\sum\Delta_z/2=2n\pi$ where $n$ is an integer, and when the $y$-component of the rotations is performed with unity fidelity i.e. $\Delta_y=0$.
Increasing the number of vertices acts to exponentially reduce the fidelity of the single logical qubit state (in this case $\mathcal{F}_N=(\mathcal{F}_1)^N,~\mathcal{F}_1\leq1$), increasingly constraining the spin control fidelity required to achieve a given spin-photon resource state fidelity.
Experimental demonstrations using single QDs have shown $\pi$-rotation fidelities as high as 99.3\%~\cite{zaporski2023ideal}.

Spin-flip errors at step 3 of the protocol has a material impact of the success probability of the fusion process.
We find the fusion process will always fail if only one of the photonic qubits input into the fusion circuit was generated in the error state $\ket{\oslash,\oslash} \pm\ket{1,1}$.
In this case fusion failure is heralded by the detection of either a single-photon or three photons.
While the single-photon detection pattern has the same signature as loss of a single photon (discussed in Sec.~\ref{sec:photon_loss}), it is important to note the two processes are not equivalent as this spin control error does not result in a wider disruption of entanglement.
Conversely, when both photonic qubits entering the fusion circuit are generated in the error state, we find that photonic fusion can successfully generate entanglement between the error states if only two photons are input into the circuit (i.e. measurement projects one input qubit into the $\ket{\oslash,\oslash}$ state while the other projected into the $\ket{1,1}$ state).
However, the detection patterns heralding successful fusion are exactly those that herald fusion failure in the absence of errors, i.e. two photons at a single detector.
Hence in addition to introducing additional failure mechanisms, spin-flip errors in the RSG protocol also introduce ambiguity into previously unambiguous detection patterns.

As discussed previously, the control processes implemented in step (5) of the protocol determine the nature of the entanglement within and between redundantly encoded vertices of the state.
Subjecting the spin flip applied in step (5a) of the protocol to $y$- and $z$-rotation errors acts to modify the internal entanglement between sub-vertices leaving the composition of individual photonic qubits unchanged (see Appendix~\ref{app:step_5a_error} for the resulting state).
$Z$-rotation errors act to introduce relative phases between early and late components of the state, while $y$-rotation errors can cause the control operation applied in step (5a) to act as either a spin-Hadamard or inverse spin-Hadamard gate rather than a spin flip.
This latter error has two main consequences.
Applying a (inverse) spin-Hadamard gate at this step of the protocol acts to effectively push sub-vertices out of the parent redundantly encoded vertex to their own vertices in the state.
Furthermore, the introduction of an unwanted (inverse) spin-Hadamard gate out of the prescribed sequence leads to a variation in the relative phase of the components of the (sub-)vertices. 
The fidelity of the resulting state accounting for this error has the same form as Eq.~\eqref{eq:fidelity_spin_flip} only now $\Delta_{y(z)}^{(n,j)}$ is the $y(z)$ error on the spin flip that moves from generating sub-vertex $j$ to sub-vertex $j+1$ further redundantly encoding the $n^{\rm th}$ logical qubit.

Errors in the the spin-Hadamard and inverse spin-Hadamard gates applied in step (5b) of the protocol have the opposite effect to that in step (5a).
Increasing the magnitude of the error increasingly causes step (5b) to act as a spin-flip which effectively pulls nominally distinct vertices in the state into a single vertex with a greater degree of redundant encoding.
\begin{figure}[t]
    \centering
    \vspace*{-0.5cm}
    \subfigim[width=7cm]{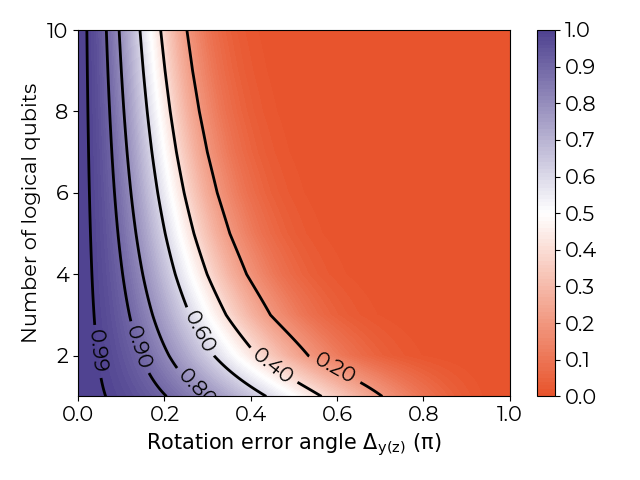}{fig:error_5bii_qubit_number_single_rotation}{72}{64}
    \vspace*{-1cm}
    \subfigim[width=7cm]{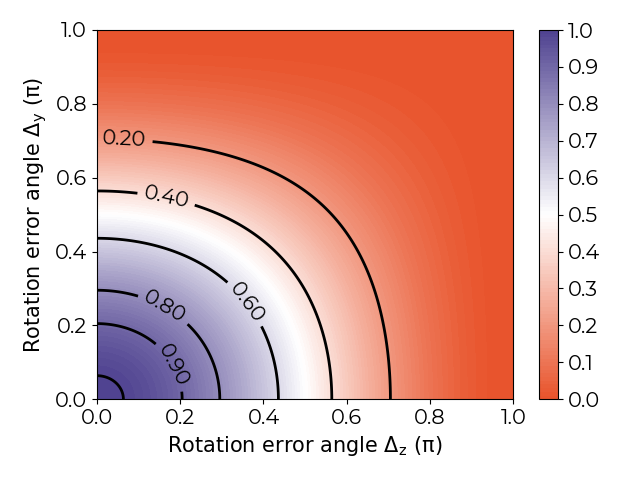}{fig:error_5bii_yz_rotation}{72}{64}
    
    \caption{The calculated resource state fidelity accounting for errors in the spin control operations in step (5bii) of the protocol. (a) The resource state fidelity as a function of a single rotation error (assuming the second rotation is performed with unity fidelity) and the number of logical qubits in the resource state. (b) The fidelity of a single redundantly encoded photonic qubit as a function of errors in the $y-$ and $z-$ rotation of the spin-qubit state.}
    \label{fig:fidelity_spin_control_error_step_5bii}
\end{figure}
The fidelity of the total generated spin-photon state in the presence of errors on the spin control gates in step (5b) of the protocol is given by
\begin{equation}
\begin{split}
    \mathcal{F}_N = \abs{\frac{1}{2^{N+1}}\Big(f_N(\Delta_y,\Delta_z) + \rm h.c.\Big)}^2
\end{split}
\end{equation}
where
\begin{equation}
    f_{a} = e^{i\Delta_z^{(n)}/2}\Big(\varepsilon_{a}(\Delta_y^{(a)})f_{a-1} + \varepsilon_{a+1}(\Delta_y^{(a)})f_{a-1}^*\Big)
\end{equation}
with $\varepsilon_x(\Delta) = \cos\Delta/2 + (-1)^x\sin\Delta/2$ and $a\geq1$.
When one of the rotations of the spin-qubit state on the block sphere is performed with unity fidelity, i.e., when either $\Delta_y = 0$ or $\Delta_z=0$, the resource state fidelity reduces to
\begin{equation}
    \mathcal{F} = \abs{\prod_n\cos\left(\frac{\Delta_j^{(n)}}{2}\right)}^2
\end{equation}
where $j\in\{y,z\}$ indicates the non-zero rotation error.
The spin-photon state fidelity accounting for spin control errors in step (5bii) is plotted in Fig.~\ref{fig:fidelity_spin_control_error_step_5bii}.

The type-II photonic fusion process is not itself necessarily directly impacted by errors on the control processes implemented in step (5) of the protocol.
When redundantly encoding photonic qubits are input into the fusion circuit, the result of the fusion operation is simply to generate entanglement between the two error states upon success.
However, the change in entanglement can result in a photonic qubit that otherwise would have redundantly encoded a larger single vertex inadvertently becoming the central photonic qubit of an additional distinct vertex in the state or vice versa.
Applying a Hadamard gate to the central qubit of a vertex will change the nature of its entanglement with neighbouring vertices.
For example, a Hadamard gate applied to the central qubit of an end vertex will in effect act to pull it into the GHZ state redundantly encoding the next vertex of the state.
Depending on the measurement outcome this can project the vertex into one of the computational basis states disrupting entanglement with neighbouring vertices.
On the other hand, directing a redundantly encoding photonic qubit that was initially designated as a central data-holding qubit to measurement will also project the entire vertex onto one of the computational basis states also disrupting entanglement with neighbouring vertices.

\subsection{Excitation-related errors}

Beyond optical control of the spin state of the QD-confined charge carrier, optical excitation of the QD system may also be subject to errors.
That is, the excitation pulses may fail to result in the emission of a photon from the selected cycling transition, or, in the case of the single-QD system, may off-resonantly excite the second optical transition co-polarised with the transition selected for photon generation resulting in the emission of photons with undesired energy.

\subsubsection{Inefficient excitation}

When excitation of an optical transition coupling a higher energy excited state $\ket{e}$ to a lower energy state is not optimised, photon emission accompanying the decay of the excited state population will only occur with a probability $p$.
In the generation of time-bin photonic qubits, probabilistic photon emission in this manner introduces a vacuum component into each of the time-bin basis states 
e.g. $\ket{\oslash,1}\rightarrow\sqrt{p}\ket{\oslash,1} + \sqrt{1-p}\ket{\oslash,\oslash}$
$\ket{\gamma}\rightarrow\sqrt{p}\ket{\gamma} + \sqrt{1-p}\ket{\oslash,\oslash}$ for $\ket{\gamma}\in\{\ket{\oslash,1},\ket{1,\oslash}\}$
(see Appendix~\ref{app:excitation_errors}).
This contrasts with the vacuum components generated by imperfect spin control in step (3) that also have a corresponding two-photon component.
As a result, the nature of the entanglement within the state remains unchanged other than the $CZ$ operators now acting on the modified basis states.
The fidelity of the resulting state, plotted in Fig~\ref{fig:off_resonant_inefficient_excitation}, takes the simple form
\begin{equation}
    \mathcal{F} = \prod_{n,m} p^{(n,m)}_\gamma
\end{equation}
where $p_\gamma^{(n,m)}$ is the probability of generating a photon when preparing the $m^{\rm th}$ qubit of the $n^{\rm th}$ vertex of the state.

When one or both of the states input into the dual-rail fusion circuit are subject to inefficient excitation errors there are two distinct potential outcomes beyond the ideal outcome.
Naturally if inefficient excitation results in no photons being generated there can be no change to the entanglement in the wider state as in effect the identity operator has been applied to the entire system.
However, if instead inefficient excitation results in only one of the photonic qubits input into the fusion circuit being generated, we find entanglement generation via photonic fusion remains possible.
In this scenario entanglement generation is heralded by detection of a single-photon at one of two out of the four detectors employed when using dual-rail qubit encoding assuming it is not known which RSG failed to generate a photon.
When both photonic qubits input into the fusion circuit were equally likely to have been generated in the vacuum state the fusion circuit can generate the same entanglement as in the absence of errors.
Detection of a photon at the other two detectors results the system being projected back into the input state minus the detected photon.

\subsubsection{Off-resonant excitation}

A further error mechanism that, from our example systems, is particularly relevant to the single charged QD system is excitation of additional detuned transitions in the system.
As pairs of transitions in the single charged QD system are co-polarised, excitation pulses will drive both transitions unless care is taken to ensure spectral overlap between the optical pulse and second (off-resonant) transition is minimised (in the QDM the off-resonant transition is also protected by polarisation). 
Assuming the co-polarised off-resonant transition is similarly perfectly cyclic, off-resonant excitation will result in the emission of a detuned $H$-polarised photon when the spin-qubit occupies the nominally dark state (spin-up in our example).
In a similar manner to inefficient excitation of the spin-qubit, the generation of undesired photon states via off-resonant excitation of additional cyclic transitions in the system modifies the time-bin basis states of the photonic qubits.
However, this now results in multi-photon components appearing in the state rather than vacuum components.
\begin{figure}[t]
    \centering
    \includegraphics[width=8cm]{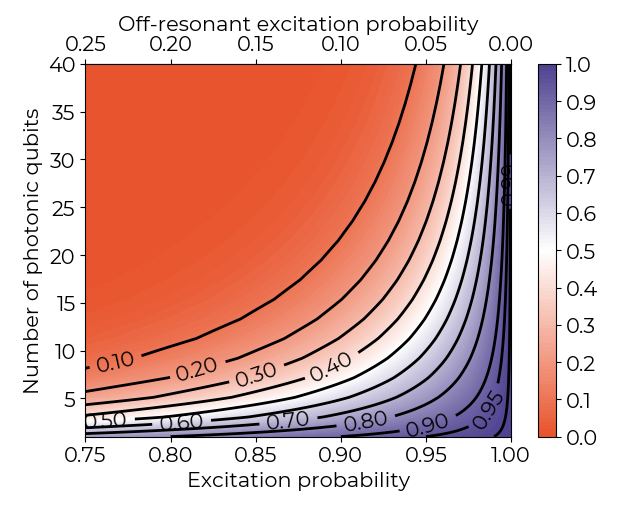}
    \caption{The calculated spin-photon state fidelity as a function of the probability of exciting the selected cyclic transition (undesired off-resonant transition) and the number of photonic qubits.}
    \label{fig:off_resonant_inefficient_excitation}
\end{figure}
The resulting state fidelity, plotted in Fig.~\ref{fig:off_resonant_inefficient_excitation}, is given by 
\begin{equation}
    \mathcal{F} = \prod_{n,m}\Big(1-p^{(n,m)}_\uparrow\Big)
\end{equation}
where $p_\uparrow^{(n,m)}$ is the probability of optically exciting the RSG when in its nominal dark state (taken to be spin-up in this example) while generating the $m^{\rm th}$ qubit of the $n^{\rm th}$ vertex of the resource state.
Assuming the undesired photon states are sufficiently spectrally distinct from the desired states, optical filtering may be employed to recover the ideal entangled spin-photon state.
However, as optical filtering is inherently probabilistic with a non-unity probability of removing undesired photon states and a non-zero probability of removing desired photon states, the inclusion of optical filtering will likely leave the resource state in a mixed state.

To understand the impact of the excitation of off-resonant transitions on the photonic fusion process we shall make two assumptions. 
First, we assume the states input into the fusion circuit were generated by identical QEs, i.e., photons emitted by corresponding transitions are indistinguishable.
We also assume that the additional photons emitted by the off-resonant transitions are completely distinguishable from the desired photon output.
Under these assumptions we find excitation of off-resonant transitions, while complicating detection patterns heralding fusion success or failure, never prevents successful entanglement generation.

When only a single additional photon is generated in the course of creating the two photonic qubits input into the fusion circuit the detection patterns heralding successful entanglement generation closely follow those in the absence of the error.
Detection of the two resonant photons at the same detector still heralds fusion failure, while the detection of these photons across the two pairs of detectors heralds fusion success.
In this way the presence of the single additional photon from one off-resonant transition is all but inconsequential in isolation, merely resulting in an additional detector click.
If instead both photonic qubits input into the fusion circuit contain an additional single-photon emitted by identical off-resonant transitions the fusion process is in effect performed twice simultaneously as the two distinguishable pairs of mutually indistinguishable photons will not interact with one another in the fusion circuit.
However, while fusion in effect occurs twice simultaneously this does not increase the probability of successful entanglement generation within a given round. This probability remains at 50\%.

\subsection{Imperfect cycling transitions}

Another error mechanism that may arise in the single-QD system occurs when the RSG does not possess a perfectly cyclic transition.
Coupling to an external photonic structure only introduces a quasi-cycling transition whose cyclicity is dependent on the strength of the polarisation-dependent relative enhancement of the optical transition decay rates~\cite{sheldon2024optical}.
Non-unity cyclicity can lead to an undesired spin flip after optical excitation for photon generation (i.e. in steps (2) and (4) of the protocol).
Not only does this leave the RSG in the incorrect spin state for further photonic qubit generation, but this event is also accompanied by the emission of a photon with undesired energy and polarisation.
The fidelity of the resulting state (given in Appendix~\ref{app:imperfect_cyclicity}) to the desired resource state takes the simple form
\begin{equation}
    \mathcal{F} = \prod_{n,m}p_{\downarrow\downarrow}^{(n,m)}
\end{equation}
where $p_{\downarrow\downarrow}^{(n,m)}$ is the probability of the RSG returning to the designated bright state (taken to be the spin-down state in this example) after optical excitation generating $m^{\rm th}$ photonic qubit of the $n^{\rm th}$ vertex of the resource state. 

When (sub-) vertices of the resource state are generated via multiple consecutive excitations the resulting resource state is particularly sensitive to this error.
The lack of spin flip operations in the generation of the GHZ (sub-)vertices when using multiple consecutive excitations results in single spin-flip errors propagating throughout the (sub-) vertex in a manner that is not seen with other error mechanisms experienced by the system.
An erroneous spin-flip will not only populate the time-bin at which the error occurred with an orthogonally polarised photon, but will cause the spin-qubit to transition into its dark state preventing the further generation of photons in the corresponding time-bins of the GHZ state (neglecting off-resonant excitation).
Moreover, if the spin-flip error occurs when generating an early time-bin component of the GHZ sub-vertex, the spin-flip error combined with the intermediate spin rotation of step (3) will result in the late time-bins of some or all (depending on whether a second erroneous spin-flip error occurs) of the photonic qubits being erroneously populated with photons of the desired frequency and polarisation.
Consequently two-photon components arise in the state whereby each time-bin of a given photonic qubit is populated with a single photon.
The lack of protection against spin-flip errors when using consecutive excitations is further highlighted when single excitation pulses are used instead.
When single excitation pulses are used in steps (2) and (4) to generate photonic qubits we find no photonic qubit remains in the vacuum state or is populated by two photons of the desired polarisation and energy in direct contract to the multi-excitation implementation.

Besides photon loss, erroneous spin-flips resulting from the imperfect cyclicity of optical transitions used for photon generation arguably have the largest impact on subsequent photonic fusion measurements.
We find the emission of an orthogonally polarised photon accompanying the spin flip will always cause the fusion operation to fail.
This failure can project one of the vertices being fused into a known state disrupting entanglement.
Furthermore, without effective discrimination between photon polarisations at the detectors or knowledge of the occurrence of an erroneous spin-flip measurement outcomes cannot distinguish between the successful generation of entanglement in the absence of the error or failure of the fusion process resulting from the error.
Mitigating against imperfect transition cyclicity requires applying polarisation selective filtering to the output of the RSGs.
However, imperfect filtering can result in a further decrease of the fusion efficiency beyond the decrease expected purely from the error.

\subsection{Photon loss}\label{sec:photon_loss}

Lastly we consider the impact of photon loss on the resulting resource state assuming the loss occurs before a Hadamard gate is applied to each of the redundantly encoding photonic qubits.
As we have encountered with other errors, optical losses modify the early and late time-bin basis states.
Explicitly tracking the loss modes, and assuming loss is uncorrelated, the resulting basis states are given by
\begin{equation}
    \begin{split}
        \ket{\oslash,1}_{n,m}\rightarrow& \sqrt{q_\oslash^{(n,m)}}\ket{\oslash,\oslash}^{L}_{n,m}\ket{\oslash,1}_{n,m} \\
        &~~~~~~~~+ \sqrt{p_\oslash^{(n,m)}}\ket{\oslash,1}^{L}_{n,m}\ket{\oslash,\oslash}_{n,m},\\
        \ket{1,\oslash}_{n,m}\rightarrow& \sqrt{\tilde{q}_\oslash^{(n,m)}}\ket{\oslash,\oslash}^{L'}_{n,m}\ket{1,\oslash}_{n,m}\\
        &~~~~~~~~+ \sqrt{\tilde{p}_\oslash^{(n,m)}}\ket{1,\oslash}^{L'}_{n,m}\ket{\oslash,\oslash}_{n,m}
    \end{split}
\end{equation}
where the probability of losing the photon from the photonic qubit indexed as $(n,m)$ is given by $p_\oslash^{(n,m)}$ and $q_\oslash^{(n,m)}=1-p_\oslash^{(n,m)}$.
The full state is given in Appendix~\ref{app:photon_loss}.
In these modified basis states we have allowed the probability of losing either an early or late photon from a given photonic qubit to be different noting that when converting to dual-rail encoding early photons must pass through a delay line which will increase the optical loss relative to photons in late time-bins.
Taking the trace over the lost photons results in the entangled spin-photon state becoming mixed, with associated classical probabilities of retaining or losing each of the photons, and completely collapses the correlations within the GHZ state when one or more photons are lost.
The fidelity of the state in the presence of photon loss takes the same form as when accounting for off-resonant excitation if the single cavity-coupled QD system
\begin{equation}
    \mathcal{F} = \prod_{n,m}q_\oslash^{(n,m)} = \prod_{n,m}\Big(1-p_\oslash^{(n,m)}\Big)_.
\end{equation}

As loss of even a single photonic qubit from a GHZ state will entirely disrupt the internal entanglement of a GHZ state, all boosted fusion operations involving any redundantly encoded vertices from which a photon was lost will fail.
For a given photon loss probability there is a threshold beyond which increasing the number of redundantly encoding photonic qubits is detrimental to the overall fusion success probability.
Assuming the redundantly encoding photonic qubits remain in the GHZ state of their parent vertices until used for a fusion operation, the probability of boosted fusion success is given by~\cite{hilaire2023near}
\begin{equation}
    P(m,\eta) = (1-2^{-m})\eta^{2m}
\end{equation}
where $m$ is the number of photonic qubits used and $1-\eta$ is the photon loss probability.
For example, at 95\% end-to-end efficiency the optimised boosted fusion approach returns a 65\% probability of successful entanglement generation vs 45\% in the standard approach with the same optical loss.

\section{Conclusion}\label{sec:conclusion}\noindent
\begin{figure}[t]
    \centering
    \includegraphics[width=1\linewidth]{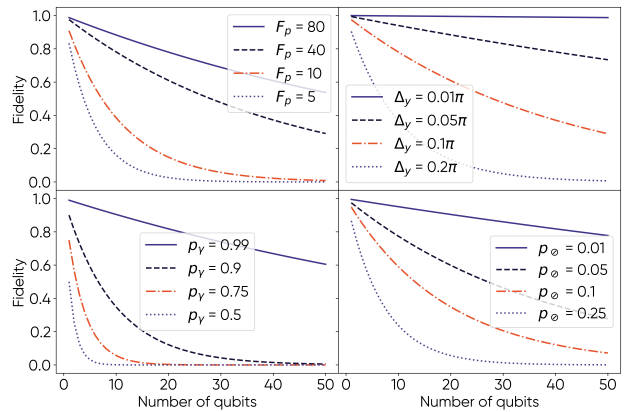}
    \caption{Calculated 1D resource state fidelity accounting for individual error mechanisms as a function of the number of generated photonic qubits. The error mechanisms considered are (top left) imperfect cyclicity $C=F_P/(F_P+1)$, (top right) imperfect spin control in step 3 of the protocol, (bottom left) inefficient photon generation, and (bottom right) photon loss.}
    \label{fig:focused_fidelity}
\end{figure}

We have proposed a general protocol for the generation of redundantly encoded photonic resource states of entangled time-bin qubits.
Our protocol is generally applicable to any quantum emitter with two ground states whose populations may be coherently transferred between one another, and possesses at least one individually-addressable cyclic transition between one of the ground states and a single excited state.
The generated photonic states enable boosted fusion to be employed for efficient generation of cluster states for MBQC or repeater states for memoryless quantum repeaters.
Furthermore, using our protocol the ordering of time-bins may be modified reducing requirements on on-chip optical routing for conversion between time-bin and dual-rail encoding of photonic qubits.
Lastly, our outlined protocol is compatible with entangled emitter schemes~\cite{economou2010optically} providing additional avenues for multi-dimension cluster state generation.

Additionally, we have individually studied how errors at each step of the protocol impact the generated resource states.
We find errors in the control and excitation of the spin-qubit are effectively teleported to the photonic component of the state and thus do not compound as an increasing error on the state of the spin-qubit as the protocol progresses.
For example, errors in the initial preparation of the spin-qubit in a superposition of the state may be negated via removal of the first generated photonic qubit from the larger resource state.
A number of other potential error mechanisms (namely spin flip errors in step (3), inefficient excitation in steps (2) and (4), and imperfect cyclic transitions) may result in individual photonic qubits being left in the vacuum state (without the destruction of entanglement associated with photon loss) or generated in a multi-photon state with a single photon per time-bin leaving the spin-qubit entangled with the photonic component in the desired superposition state.
Moreover, control errors in the protocol not only impact the composition of individual photonic qubits, but may also change the entanglement between individual photonic qubits and between vertices in the resource states.
Each of these errors impacts subsequent photonic fusion processes.
Depending on the error this can manifest as an ambiguity in the expected detector patterns or cause failure of entanglement generation.

Fig.~\ref{fig:focused_fidelity} highlights calculated state fidelities accounting for individual error mechanisms (imperfect cyclicity, spin control errors, inefficient excitation, and photon loss) as a function of the size of the generated resource state.
Here we consider a range of performance values encompassing experimentally achievable, state-of-the-art, and beyond state-of-the-art.
However, we note that while the state fidelity is a useful comparative measure for studying the relative impact of error mechanisms or comparing emitter performance, it does not necessarily translate directly to the overall performance of quantum photonic technologies employing resource state generators.
While the generation of resource states may form the basis of a fault-tolerant computing or repeater architecture, the question of the optimal route to constructing such architectures remains an open one.
Answering this question will require combining the impact of the discussed error mechanisms, along with others such as imperfect single-photon purity indistinguishability and emitter coherence, into a single efficient error model that can be used to accurately inform quantum error correction models of physical errors.
Only when considering this built-in error tolerance can system performance actually be deduced.

\bibliography{ref}

\appendix

\onecolumngrid

\section{Polarisation / Dual-rail encoded resource state generation protocol}\label{app:dual_rail_encoded_resource_state}

The generation of small redundantly encoded resource states of polarisation encoded photonic resource states using a singly charged QD coupled to a micropillar cavity in a weak out-of-plane magnetic field has been demonstrated in~\cite{huet2025deterministic}.
Furthermore, in~\cite{vezvaee2022deterministic} the authors note that QDMs could also be used for generating polarisation encoded photonic qubits if the energies of the light emitted by the direct optical transitions can be brought into resonance.
While on-chip optical waveguides do not generally preserve photon polarisation, disqualifying the use of polarisation encoded photonic qubits in such structures, situating either a source of entangled polarisation encoded photonic qubits in an on-chip directional photonic structure~\cite{mahmoodian2016engineering} would enable the direct on-chip generation of dual-rail photonic qubits.
Our protocol only requires simple modification to be compatible with resource state generators of this type:
\begin{enumerate}
    \item 
    \begin{enumerate}
        \item Initialise the spin state of the spin-qubit into one of the charge-carrier basis states $\{\ket{\downarrow}, \ket{\uparrow}\}$,
        \item Apply an (inverse) spin-Hadamard gate to prepare a the charge carrier in a 50:50 superposition of its spin basis states,
    \end{enumerate}
    \item Simultaneously excite the two direct transitions of the spin-qubit with $M_i$ pulses to redundantly encode a single logical qubit on $M_i$ dual-rail photonic qubits,
    \item Either:
    \begin{enumerate}
        \item Apply a $2\pi$ spin rotation to continue redundantly encoding the logical qubit on additional dual-rail photonic qubits, or
        \item
        \begin{enumerate}
            \item if, in step (1b), the spin-qubit was prepared in the $\ket{-}$ ($\ket{+}$) state, apply (inverse) spin-Hadamard gates in subsequent cycles of the protocol, or
            \item alternately apply spin-Hadamard and inverse spin-Hadamard gates in subsequent cycles of the protocol if the spin-qubit was prepared in the $\ket{+}$ state in step (1b), reversing the order if prepared in the $\ket{-}$ state.
        \end{enumerate}
    \end{enumerate}
    \item Repeat steps (2) and (3) of the protocol $N$ times to generate a redundantly encoded photonic resource state with $N$ logical qubits.
\end{enumerate}
Directly generating polarisation / dual-rail photonic qubits simplifies the resource state generation protocol, removing the need for a spin-qubit flip and second excitation which in turn may reduce the number of errors, and also removes the need for the delay lines to convert between time-bin and dual-rail encoding reducing optical loss.
Moreover, we find imperfect directionality of the resulting photon emission when generating a given vertex of the state results in $Z$ errors being applied to its neighbouring vertices which quantum error correction codes are well suited to detecting and correcting.
Naturally one would then need to account for the opposite circular polarisations of the photons in the two different waveguide modes when performing photonic fusion.

\section{Spin-qubit rotation operators}

The spin Hadamard $H_s$ and inverse spin Hadamard $\bar{H}_s$ gates act on the spin basis states as
\begin{equation}
\begin{split}
    H_s\ket{\uparrow} =& \frac{1}{\sqrt{2}}(\ket{\downarrow}+\ket{\uparrow}),~H_s\ket{\downarrow} = \frac{1}{\sqrt{2}}(\ket{\downarrow}-\ket{\uparrow}),\\
    \bar{H}_s\ket{\uparrow} =& \frac{1}{\sqrt{2}}(\ket{\downarrow}-\ket{\uparrow}),~\bar{H}_s\ket{\downarrow} = \frac{-1}{\sqrt{2}}(\ket{\downarrow}+\ket{\uparrow}).
\end{split}
\end{equation}

To account for errors in the rotation of the spin-qubit state we modify the operators such that a spin-Hadamard gate with errors is described by
\begin{equation}
\begin{split}
    H_s(\Delta_y,\Delta_z) = R_z(\Delta_z)&R_y(\pi/2 + \Delta_y) =
    \frac{1}{\sqrt{2}}\begin{pmatrix}
    e^{-i\Delta_z/2}\varepsilon_-(\Delta_y) & -e^{-i\Delta_z/2}\varepsilon_+(\Delta_y)\\ e^{i\Delta_z/2}\varepsilon_+(\Delta_y) & e^{i\Delta_z/2}\varepsilon_-(\Delta_y)
    \end{pmatrix}
\end{split}
\end{equation}
while an inverse spin-Hadamard gate with errors is given by
\begin{equation}
\begin{split}
    \bar{H}_s(\Delta_y,\Delta_z) = R_z(\Delta_z)R_y(3\pi/2 + \Delta_y) =
    \frac{1}{\sqrt{2}}\begin{pmatrix}
    -e^{-i\Delta_z/2}\varepsilon_+(\Delta_y) & -e^{-i\Delta_z/2}\varepsilon_-(\Delta_y)\\ e^{i\Delta_z/2}\varepsilon_-(\Delta_y) & -e^{i\Delta_z/2}\varepsilon_+(\Delta_y)
    \end{pmatrix}
\end{split}
\end{equation}
where $\varepsilon_\pm(\Delta) = \cos\Delta/2 \pm \sin\Delta/2$.
The spin flip operator with errors is given by
\begin{equation}
    \begin{split}
        R_z(\Delta_z)&R_y(\pi + \Delta_y)=
        \begin{pmatrix}
            -e^{-i\Delta_z/2}\sin{\Delta_y/2} & -e^{-i\Delta_z/2}\cos{\Delta_y/2}\\
            e^{i\Delta_z/2}\cos{\Delta_y/2} & -e^{i\Delta_z/2}\sin{\Delta_y/2}
        \end{pmatrix}.
    \end{split}
\end{equation}

\section{Type-II photonic fusion}\label{app:fusion}

\begin{figure}
    \centering
    \includegraphics[width=8cm]{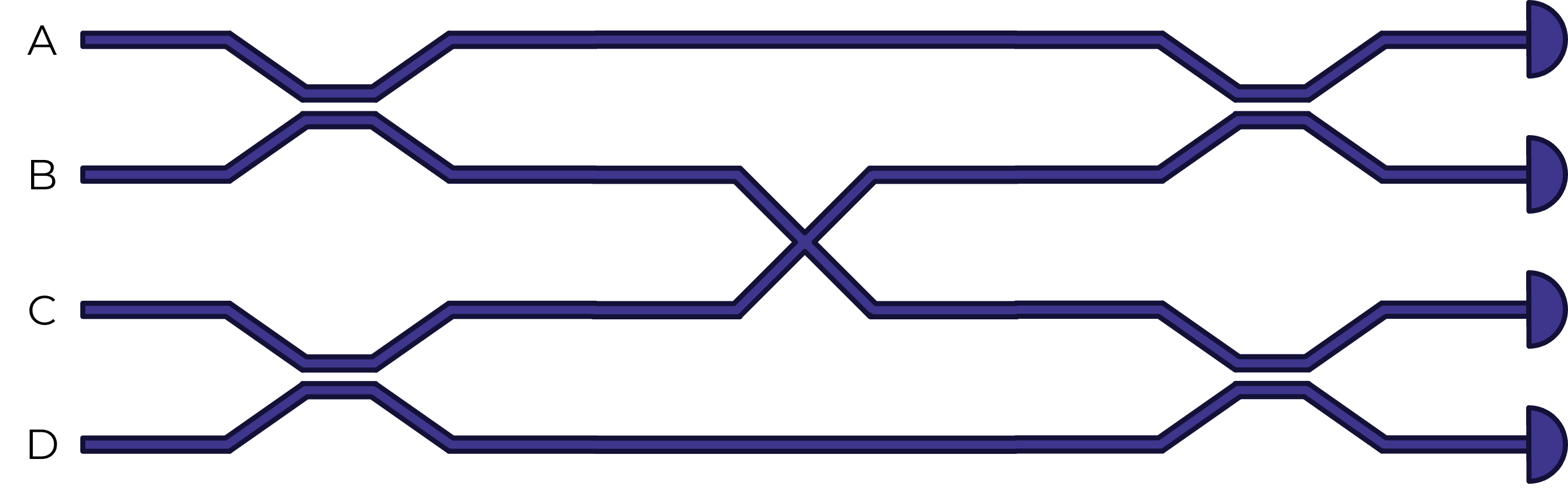}
    \caption{An example schematic of a photonic integrated circuit that performs a fusion operation between redundantly encoded resource states of dual-rail encoded photonic qubits. The circuit starts with a pair of 50:50 waveguide beam splitters acting between modes A and B, and modes C and D. After these initial beam splitters modes B and C are swapped before being input into a second pair of 50:50 beam splitters. At the output of the circuit four single-photon detectors measure the resulting state. Components may be added or removed from the circuit to change the variant of type-II fusion being performed, e.g. XXZZ, XZZX etc..}
    \label{fig:fusion_circuit}
\end{figure}

The aim of the presented protocol is to generate entangled photon states that enable the probability successful entanglement generation via photonic fusion to be boosted from 50\% in standard guise to near unity via a time-delayed repeat-until-success strategy.
Here we present a brief review of the type-II photonic fusion process with redundantly encoded resource states.

To generate entanglement between two redundantly encoded vertices, either in the same state or in different states, we must first convert from time-bin to dual-rail qubit encoding.
In our description of the state this simply requires relabelling the time-bin states as spatial modes, and in practice may be achieved with a photonic switch and a delay line.
We can then rewrite the ideal redundantly encoded resource state in terms of photon creation operators for the four input modes of the type-II photonic fusion circuit shown in Fig.~\ref{fig:fusion_circuit}
\begin{equation}
    \ket{\psi_{\rm in}} = \Big(f_AA_{\rm in}^\dagger + f_B B_{\rm in}^\dagger\Big)\otimes(f_CC_{\rm in}^\dagger + f_DD_{\rm in}^\dagger\Big)\ket{\oslash}
    \label{eq:fusion_in}
\end{equation}
where $f_X$ for $X\in\{A,B,C,D\}$ are functions of creation operators $X^\dagger$ that generate the wider redundantly encoded resource states to be fused.
Assuming a lossless circuit and that the input photons are indistinguishable, we can construct a transfer matrix relating the photon creation operators at the outputs of the circuit to the photon creation operators at the inputs
\begin{equation}
    \begin{pmatrix}
        A_{\rm out}^\dagger \\ B_{\rm out}^\dagger \\ C_{\rm out}^\dagger \\ D_{\rm out}^\dagger
    \end{pmatrix}
    = \frac{1}{2}
    \begin{pmatrix}
        1 & 1 & 1 & 1 \\ 1 & 1 & -1 & -1 \\ 1 & -1 & 1 & -1 \\ 1 & -1 & -1 & 1
    \end{pmatrix}
    \begin{pmatrix}
        A_{\rm in}^\dagger \\ B_{\rm in}^\dagger \\ C_{\rm in}^\dagger \\ D_{\rm in}^\dagger
    \end{pmatrix}.
    \label{eq:type_ii_mod}
\end{equation}
Using this relationship and the input state, the state at the output of the circuit is given by
\begin{equation}
    \begin{split}
        \ket{\psi_{\rm out}} =& \frac{1}{8}\Big\{(f_A+f_B)(f_C+f_D)((A_{\rm out}^\dagger)^2 - ({B_{\rm out}}^\dagger)^2)
        +(f_A-f_B)(f_C-f_D)(({C_{\rm out}}^\dagger)^2 - ({D_{\rm out}}^\dagger)^2)\\
        &+2(f_Af_C - f_Bf_D)(A_{\rm out}^\dagger {C_{\rm out}}^\dagger + {B_{\rm out}}^\dagger {D_{\rm out}}^\dagger)
        +2(f_Af_D - f_Bf_C)(A_{\rm out}^\dagger {D_{\rm out}}^\dagger + {B_{\rm out}}^\dagger {C_{\rm out}}^\dagger)\Big\}\ket{\oslash}.
    \end{split}
\end{equation}
Hence when both photons input into the circuit are detected at the same detector ($(X_{\rm out}^\dagger)^2$) the two resource states remain in the separable state $(f_A\pm f_B)\otimes(f_C\pm f_D)\ket{\oslash}$.
However, detection of the two photons at one detector of each of the pair of dual-rails projects the previously separable state to one of two entangled states, $(f_Af_C - f_Bf_D)\ket{\oslash}$ or $(f_Af_D - f_Bf_C)\ket{\oslash}$, depending on the detector pattern.
The probability of projecting the input state into either of these two entangled states is 50\% and thus there is also a 50\% probability of projecting the input states into one of the separable states.

\section{Protocol errors}

In this appendix we provide the full forms of the resulting resource states when the protocol is subject to errors at its different steps.

\subsection{Step(1)}\label{app:initialisation_error}

The first errors we consider concern the preparation of the spin qubit before generation of the photonic state.
In the main text we consider the case where both the chosen initialisation protocol prepares the spin qubit in an arbitrary mixed state $(\abs{\alpha}^2\ketbra{\downarrow}{\downarrow} + \abs{\beta}^2\ketbra{\uparrow}{\uparrow})$ with $\abs{\alpha}^2 + \abs{\beta}^2=1$ rather than in a desired pure state and the subsequent spin control fails to prepare the desired spin superposition state.
As a result of these errors the generated state is given by
\begin{equation}
    \sigma_{\rm RE} = \mathcal{F}_{s}\ketbra{\Phi_{\rm RE}}{\Phi_{\rm RE}} + (1-\mathcal{F}_{s})\ketbra{\Phi'_{\rm RE}}{\Phi'_{\rm RE}}
\end{equation}
where $\mathcal{F}_{s} = \bra{s}\frac{1}{2}(\abs{\alpha}^2\ketbra{\downarrow}{\downarrow} + \abs{\beta}^2\ketbra{\uparrow}{\uparrow})\ket{s}$  is the fidelity of the prepared spin state to the target spin state $s\in\{\uparrow,\downarrow\}$,
\begin{equation}
    \begin{split}
        \ket{\Phi_{\rm RE}} =&~ \mathbf{CZ_{\uparrow,\mathcal{V}_N}}\Bigg(\prod_{n=1}^{N-1}\mathbf{CZ}_{\mathcal{V}_{n+1},\mathcal{V}_n}\Bigg)
        \mathbf{Z}_\uparrow^N\ket{\pm}_S\ket{+}_{n,M_n}^{\otimes_{N-1}}
        \frac{e^{\frac{i\Delta_z}{2}}}{\sqrt{2}}\Bigg[\varepsilon_\pm(\Delta_y)\ket{E}_1^{\otimes_{M_1}} + e^{-i\Delta_z}\varepsilon_\mp(\Delta_y)\ket{L}_1^{\otimes_{M_1}}\Bigg],
    \end{split}
\end{equation}
and
\begin{equation}
    \begin{split}
        \ket{\Phi_{\rm RE}'} =&~ \mathbf{CZ_{\uparrow,\mathcal{V}_N}}\Bigg(\prod_{n=1}^{N-1}\mathbf{CZ}_{\mathcal{V}_{n+1},\mathcal{V}_n}\Bigg)
        \mathbf{Z}_\uparrow^N\ket{\pm}_S\ket{+}_{n,M_n}^{\otimes_{N-1}}
        \frac{e^{\frac{i\Delta_z}{2}}}{\sqrt{2}}\Bigg[\varepsilon_\mp(\Delta_y)\ket{E}^{\otimes_{M_1}} - e^{-i\Delta_z}\varepsilon_\pm(\Delta_y)\ket{L}^{\otimes_{M_1}}\Bigg],
    \end{split}
\end{equation}
where $\mathbf{Z}_\uparrow$ applies a phase of $-1$ to the spin-up component of the spin state $\ket{\pm}_S=(\ket{\downarrow}\pm\ket{\uparrow})/\sqrt{2}$, $\ket{E}^{\otimes_{M_n}}=\bigotimes_1^{M_n}\ket{\oslash,1}_{n,{M_n}}$ is the early component of the photonic vertices, and $\ket{L}^{\otimes_{M_n}}=\bigotimes_1^{M_n}\ket{1,\oslash}_{n,{M_n}}$ is the late component of the photonic vertices.

In the main text we also state that the spin qubit may be prepared in a pure state from an arbitrary mixed state via generation and measurement of a single photonic qubit.
Starting from the mixed state $\rho_0 = \abs{\alpha}^2\ketbra{\downarrow}{\downarrow} + \abs{\beta}^2\ketbra{\uparrow}{\uparrow}$, following steps 2-4 of the protocol will result in the generation of the state
\begin{equation}
    \rho = \abs{\alpha}^2\ketbra{\downarrow}{\downarrow}_s\otimes\ketbra{\oslash,1}{\oslash,1}_\gamma + \abs{\beta}^2\ketbra{\uparrow}{\uparrow}_s\otimes\ketbra{1,\oslash}{1,\oslash}_\gamma.
\end{equation}
Rather than applying a spin-Hadamard operation, we can instead perform a measurement on the photonic qubit. 
As this photonic qubit is generated in a Bell state with the spin-qubit, this measurement will project the spin qubit into a pure state $\rho_s=\Tr_\gamma(M\rho)$ where $M\in\{\ketbra{\oslash,1}{\oslash,1},\ketbra{1,\oslash}{1,\oslash}\}$.

\subsection{Step (3)}\label{app:step_3_error}

Step (3) of the protocol is responsible for ensuring each of the generated photonic qubits has the desired composition and phase relation.
As a result, any errors in this step of the protocol will act to modify the composition of the generated photonic qubits and modify the relative phase between their components.
Accounting for this error the resulting entangled spin-photon state may be written as
\begin{equation}
\begin{split}
    \ket{\Phi_{\rm RE}}_N =&
    \mathbf{CZ}_{\uparrow,\mathcal{V}_{N}}^\prime
    \Bigg(\prod_{n=1}^{N-1}\mathbf{CZ}_{\mathcal{V}_{n+1}, \mathcal{V}_n}^\prime\Bigg)
    Z_\uparrow^N\ket{\pm}_S
    \bigotimes_{n=1}^N\frac{1}{\sqrt{2}}\Bigg\{\Bigg(\prod_{j=1}^{J_n-1}\mathbf{\mathcal{E}}_{\mathcal{S}_{n,j+1},\mathcal{S}_{n,j}}^\prime\Bigg)\\
    &\bigotimes_{j=1}^{J_n}\Bigg(\cos{\frac{\Delta_y^{(n,j)}}{2}}\Big(e^{-i\frac{\Delta_z^{(n,j)}}{2}}\bigotimes_{m\in \mathcal{S}_{n,j}}\ket{\oslash,1}_{n,m}
    + e^{i\frac{\Delta_z^{(n,j)}}{2}}\bigotimes_{m\in \mathcal{S}_{n,j}}\ket{1,\oslash}_{n,m}\Big)\\
    &~~~~~~~~\mp (-1)^n\sin{\frac{\Delta_y^{(n,j)}}{2}}\Big(e^{-i\frac{\Delta_z^{(n,j)}}{2}}\bigotimes_{m\in \mathcal{S}_{n,j}}\ket{\oslash,\oslash}_{n,m}
    - e^{i\frac{\Delta_z^{(n,j)}}{2}}\bigotimes_{m\in \mathcal{S}_{n,j}}\ket{1,1}_{n,m}\Big)\Bigg)\Bigg\}
\end{split}
\label{eq:state_vector_flip_generation_error}
\end{equation}
where $\Delta_y$ and $\Delta_z$ are errors on the $y$- and $z$-rotations of the spin state respectively.
Modification of the photonic qubit composition accordingly modifies the operators generating entanglement between $CZ'$ and within $\mathcal{E}'$ the redundantly encoded vertices of the state.

Neglecting the phase error resulting from an imperfect $z$-rotation, following Appendix~\ref{app:fusion} the impact of this error on the fusion process can be understood by rewriting the error state in terms of photon creation operators 
\begin{equation}
\begin{split}
    \ket{\psi_{\rm in}} =& \frac{1}{2}\Bigg(\cos{\frac{\Delta_y^{(1)}}{2}}(f_A{A}_{\rm in}^\dagger + f_B{B}_{\rm in}^\dagger)\pm\sin{\frac{\Delta_y^{(1)}}{2}}(\bar{f}_A\mathbf{I}_{AB} - \bar{f}_B{A}_{\rm in}^\dagger {B}_{\rm in}^\dagger)\Bigg)\\
    &\otimes\Bigg(\cos{\frac{\Delta_y^{(2)}}{2}}(f_C{C}_{\rm in}^\dagger + f_D{D}_{\rm in}^\dagger)\pm\sin{\frac{\Delta_y^{(2)}}{2}}(\bar{f}_C\mathbf{I}_{CD} - \bar{f}_D{C}_{\rm in}^\dagger {D}_{\rm in}^\dagger)\Bigg)\ket{\oslash}
\end{split}
\end{equation}
where $\bar{f}$ are functions of creation operators generating the wider state when an error occurred in the generation of the photonic qubit being used for fusion.
Using the Eq.~\eqref{eq:type_ii_mod} the state output by the type-II photonic fusion circuit shown in Fig.~\ref{fig:fusion_circuit} can be calculated.
There are four different outcomes relating to the four different input scenarios.
When the qubit generation process occurred without error for both photonic qubits input into the circuit fusion proceeds as discussed in Appendix~\ref{app:fusion}.
If instead an error occurred when generating one of the input photonic qubits but not the other fusion fails indicated by the detection of a single photon or three photons.
However, when the generation of both photonic qubits input into the fusion circuit is subject to a spin flip error at step 3 of the protocol entanglement generation between the error may still occur. 
This is heralded by the detection of two photons, either at the same detector or across the pair of detectors at the output of a pair of spatial modes defining a dual-rail qubit.

\subsection{Step (5a)}\label{app:step_5a_error}

The choice of control operation in step (5) of the protocol determines the nature of the entanglement between the previously generated photonic qubit(s) and the photonic qubit(s) generated in the next cycle of the protocol.
This step is further subdivided into two sub-steps.
The operations outline in step (5a) prepare the RSG to further redundantly encode the  vertex of the resource state currently being generated.
As such, errors at this step of the protocol act to modify the internal entanglement of the vertex such that the resulting state is given by
\begin{equation}
\begin{split}
    \ket{\Phi_{\rm RE}}_{N} =&~ \mathbf{CZ}_{\uparrow,\mathcal{V}_N}\Bigg(\prod_{n=1}^{N-1} \mathbf{CZ}_{\mathcal{V}_n,\mathcal{V}_{n+1}}\Bigg) Z_\uparrow^N\ket{\pm}_S
    \bigotimes_{n=1}^N\frac{1}{\sqrt{2}}\Bigg\{\Bigg(\prod_{j=1}^{J_N-1} \mathbf{\bar{\mathcal{E}}}_{\mathcal{S}_{n,j+1},\mathcal{S}_{n,j}}\Big(\Delta_y^{(n,j)},\Delta_z^{(n,j)}\Big)\Bigg)\\
    &\bigotimes_{j=1}^{J_n}\Bigg(\bigotimes_{m\in \mathcal{S}_{n,j}}\ket{\oslash,1}_{n,m} + \bigotimes_{m\in \mathcal{S}_{n,j}}\ket{1,\oslash}_{n,m}\Bigg)\Bigg\}.
\end{split}
\label{eq:state_vector_spin_flip_ghz_entanglement_error}
\end{equation}
where $\bar{\mathcal{E}}_{\mathcal{S}_{n,j+1},\mathcal{S}_{n,j}}\Big(\Delta_y^{(n,j)},\Delta_z^{(n,j)}\Big)$ is the modified operator determining the entanglement within a given vertex of the photonic component of the state.
As this error does not change the composition of the individual photonic qubits, the state input into the fusion circuit maintains the same form other than the $f_X$ operators being changed to reflect the different entanglement structure.

\subsection{Step (5b)}\label{app:step_5b_error}

In contrast to step (5a), step (5b) of the protocol prepares the RSG to generate an additional vertex in the resource state.
When the protocol is subject to spin control errors at this step it is thus the entanglement between vertices in the resource state that is impacted.
These spin control errors yield the state
\begin{equation}
\begin{split}
    \ket{\Phi_{\rm RE}}_{N} =&~ \mathbf{CZ}_{\uparrow,\mathcal{V}_N}(\Delta_y^{(N)},\Delta_z^{(N)})
    \Bigg(\prod_{n}^{N-1} \mathbf{CZ}_{\mathcal{V}_{n+1},\mathcal{V}_{n}}\Big(\Delta_{y}^{(n)},\Delta_z^{(n)}\Big)\Bigg)
    Z_\uparrow^N\ket{\pm}_S
    \bigotimes_{n=1}^N\ket{+}_{n,M_n}
\end{split}
\label{eq:state_vector_spin_had_error}
\end{equation}
with modified entanglement between vertices accounted for by the $\mathbf{CZ}(\Delta_y,\Delta_z)$ operators.
Similar to errors in step (5a), as this error does not change the composition of the individual photonic qubits the state input into the fusion circuit maintains the same form other than the $f_X$ operators being changed to reflect the different entanglement structure.

\subsection{Imperfect Cycling Transitions}\label{app:imperfect_cyclicity}

One of the more significant errors occurs when the system acting as the resource state generator does not possess perfectly cyclic transitions, but rather quasi-cyclic transitions.
Taking the example of the single charged QD in a magnetic field, the lack of true cyclic transitions results in a non-zero probability of unwanted spin-flips occurring during the excitation steps of the protocol accompanied by the emission of spectrally distinct photons with orthogonal polarisation to the desired photon output.
When (sub-)vertices of the resource state are generated via multiple consecutive excitations in steps (2) and (4) the early and late basis states of the individual photonic qubits are modified by the spin-flip error to
\begin{equation}
    \begin{split}
        \ket{E}_{n,j\in\mathcal{V}_n} &= \Bigg\{\Bigg(\bigotimes_{m\in\mathcal{S}_{n,j}}\sqrt{p_{\downarrow\downarrow}^{(n,m)}}\ket{\oslash,\Tilde{H}}_{n,m}\Bigg)\\ 
        &- \sum_{m\in{\mathcal{S}_{n,j}}}\Bigg[\Bigg(\bigotimes_{b=m+1}^{\max\mathcal{S}_{n,j}}\ket{\oslash,\oslash}_{n,b}\Bigg)\otimes\sqrt{\bar{q}_{\downarrow\downarrow}^{(m)}q_{\downarrow\downarrow}^{(m)}}\ket{V,V}_{n,m}\otimes\Bigg(\bigotimes_{a=\min\mathcal{S}_{n,j}}^{m-1>0}\sqrt{\bar{p}_{\downarrow\downarrow}^{(a)}p_{\downarrow\downarrow}^{(a)}}\ket{\Tilde{H},\Tilde{H}}_{n,a}\Bigg)\Bigg]\\
        &-\sum_{m,m+k\in\mathcal{S}_{n,j}}\Bigg[\Bigg(\bigotimes_{c=m+k+1}^{\max S_{n,j}}\ket{\oslash,\oslash}_{n,c}\Bigg)\otimes\sqrt{\bar{p}_{\downarrow\downarrow}^{(m+k)}}\ket{V,\oslash}_{n,m+k}\otimes\Bigg(\bigotimes_{b=m+1}^{m+k-1}\sqrt{p_{\downarrow\downarrow}^{(n,b)}}\ket{\Tilde{H},\oslash}_{n,b}\Bigg)\\
        &~~~~~~~~~~~~~~~~~~~~~~~~~~~~\otimes\sqrt{\bar{p}_{\downarrow\downarrow}^{(n,m)}}\sqrt{q_{\downarrow\downarrow}^{(n,m)}}\ket{\Tilde{H},V}_{n,m}\otimes\Bigg(\bigotimes_{a=\min\mathcal{S}_{n,j}}^{m-1}\sqrt{\bar{p}_{\downarrow\downarrow}^{(n,a)}p_{\downarrow\downarrow}^{(n,a)}}\ket{\Tilde{H},\Tilde{H}}_{n,a}\Bigg)\Bigg]\\
        &-\sum_{m,m+k\in\mathcal{S}_{n,j}}\Bigg[\Bigg(\bigotimes_{c=m+1}^{\max\mathcal{S}_{n,j}}\ket{\oslash,\oslash}_{n,c}\Bigg)\otimes\sqrt{p_{\downarrow\downarrow}^{(n,m)}}\ket{\oslash,V}_{n,m}\otimes\Bigg(\bigotimes_{b=m-k+1}^{m-1}\sqrt{p_{\downarrow\downarrow}^{(n,b)}}\ket{\oslash,\Tilde{H}}_{n,b}\Bigg)\\
        &~~~~~~~~~~~~~~~~~~~~~~~~~~\otimes\sqrt{\bar{p}_{\downarrow\downarrow}^{(m-k)}}\sqrt{q_{\downarrow\downarrow}^{(m-k)}}\ket{V,\Tilde{H}}_{m-k}\otimes\Bigg(\bigotimes_{a=\min\mathcal{S}_{n,j}}^{i-k-1}\sqrt{\bar{p}_{\downarrow\downarrow}^{(n,a)}p_{\downarrow\downarrow}^{(n,a)}}\ket{\Tilde{H},\Tilde{H}}_{n,a}\Bigg)\Bigg]\\
        &-\sum_{m\in\mathcal{S}_{n,j}}\Bigg[\Bigg(\bigotimes_{b=m+1}^{\max\mathcal{S}_{n,j}}\ket{\oslash,\oslash}_{n,b}\Bigg)\otimes\sqrt{\bar{p}_{\downarrow\downarrow}^{(n,m)}}\ket{V,\oslash}_{n,m}\otimes\Bigg(\bigotimes_{a=\min\mathcal{S}_{n,j}}^{m-1}\sqrt{p_{\downarrow\downarrow}^{(n,a)}}\ket{\Tilde{H},\oslash}_{n,a}\Bigg)\Bigg]\Bigg\},\\
        \ket{L}_{n,j\in\mathcal{V}_n} &= \Bigg\{\Bigg(\bigotimes_{m\in\mathcal{S}_{n,j}}\sqrt{\bar{p}_{\downarrow,\downarrow}^{(n,m)}}\ket{\Tilde{H},\oslash}_{n,m}\Bigg)
        - \sum_{m\in\mathcal{S}_{n,j}}\Bigg[\Bigg(\bigotimes_{b=m+1}^{\max\mathcal{S}_{n,j}}\sqrt{p_{\downarrow\downarrow}^{(n,b)}}\ket{\Tilde{H},\oslash}_{n,b}\Bigg)\\
        &~~\otimes\sqrt{\bar{p}_{\downarrow\downarrow}^{(n,m)}q_{\downarrow\downarrow}^{(n,m)}}\ket{\Tilde{H},V}_{n,m}
        \otimes\Bigg(\bigotimes_{n=1}^{i-1}\sqrt{\bar{p}_{\downarrow\downarrow}^{(n,a)}p_{\downarrow\downarrow}^{(n,a)}}\ket{\Tilde{H},\Tilde{H}}_{n,a}\Bigg)\Bigg]\Bigg\}.
    \end{split}
\end{equation}
Here we change notation from photon numbers to polarisation ($H$ and $V$) and energy (indicated by a tilde).
The first term in $\ket{E}$ accounts for no errors when generating the early time-bin components, the second when two spin-flip errors occur at the same time-bin photonic qubit, the third when a spin-flip error occurs at qubit $m$ in the first excitation step and qubit $m+k$ in the second excitation step, the fourth when a spin-flip error occurs at qubit $m$ in the first excitation step and qubit $m-k$ in the second excitation step, and the fifth when an error occurs at qubit $m$ when generating photonic qubits in the late time-bin state.
In the $\ket{L}$ component of the state the first term accounts for no errors when generating the late components of the photonic qubits while the second term accounts for a spin-flip error occurring at qubit $m$ in the first excitation step while no error occurs in the second excitation step.

Our protocol is particularly sensitive to the lack of perfectly cyclic transitions when multiple consecutive excitation pulses are used to generate the (sub-)vertices of the resource state.
The reduced number of control operations allows this error to permeate through larger portions of the generated state than found with other errors.
This lack of protection against spin-flip errors when using consecutive excitations is further highlighted when single excitation pulses are used instead.
The modification of the early and late components of the redundantly encoded logical qubits are simplified to
\begin{equation}
    \begin{split}
        \ket{E}_{n,m} =& \sqrt{p_{\downarrow\downarrow}^{(n,m)}}\ket{\oslash,\tilde{H}}_{n,m}
        - \sqrt{\bar{q}_{\downarrow\downarrow}^{(n,m)}q_{\downarrow\downarrow}^{(n,m)}}\ket{V,V}_{n,m}
        - \sqrt{\bar{q}_{\downarrow\downarrow}^{(n,m)}}\ket{V,\oslash}_{n,m}\\
        \ket{L}_{n,m} =&\sqrt{\bar{p}_{\downarrow\downarrow}^{(n,m)}}\ket{\tilde{H},\oslash}_{n,m}
        + \sqrt{\bar{p}_{\downarrow\downarrow}^{(n,m)}q_{\downarrow\downarrow}^{(n,m)}}\ket{\tilde{H},V}_{n,m}.
    \end{split}
\end{equation}
From this simplified state we see that no photonic qubit remains in the vacuum state, or is populated by two photons with the desired polarisation and energy unlike when consecutive excitation is employed.

Considering performing photonic fusion between two redundantly encoded vertices, the input state can be expressed as
\begin{equation}
\begin{split}
    \ket{\psi_{\rm in}} =& \Big\{ f_A(\sqrt{p}A^\dagger_H - \sqrt{q}B_V^\dagger - \sqrt{qq}A^\dagger_VB^\dagger_V) + f_B(\sqrt{p}B^\dagger_H + \sqrt{pq}A^\dagger_VB^\dagger_H)\Big\}\\
    &\otimes\Big\{ f_C(\sqrt{p}C^\dagger_H - \sqrt{q}D_V^\dagger - \sqrt{qq}C^\dagger_VD^\dagger_V) + f_D(\sqrt{p}D^\dagger_H + \sqrt{pq}C^\dagger_VD^\dagger_H)\Big\}
\end{split}
\end{equation}
where the subscripts $H$ and $V$ indicate the photon polarisation.
Performing the transformation reveals the output state has 2, 3, and 4 photon components and only when no spin flips have occurred can the remaining state be projected into an entangled state.

\subsection{Steps (2) and (4)}\label{app:excitation_errors}

Steps (2) and (4) of the protocol are responsible for the generation of the single-photons that form the fundamental physical component of the generated photonic qubits. 
Assuming the RSG acts as a source of pure single photons, steps (2) and (4) can be subject to two errors, inefficient excitation of the selected transition and excitation of transitions other than that selected for photon generation.

When excitation of the RSG fails to result in the emission of a photon the resulting resource state is given by
\begin{equation}
    \begin{split}
        \ket{\Phi_{\rm RE}}_N = \mathbf{CZ}_{\uparrow,\mathcal{V}_N}\Bigg(\prod_{n=1}^{N-1}\mathbf{CZ}_{\mathcal{V}_{n+1},\mathcal{V}_n}\Bigg)Z_\uparrow^N\ket{\pm}_S
        \bigotimes_{n=1}^N\frac{1}{\sqrt{2}}&\Bigg\{\bigotimes_{m=1}^{M_n}\Big(\sqrt{1-p^{(n,m)}_\gamma}\ket{\oslash,\oslash}_{n,m}
        + \sqrt{p^{(n,m)}_\gamma}\ket{\oslash,1}_{n,m}\Big)\\
        &+ \bigotimes_{m=1}^{M_n}\Big(\sqrt{1-p^{(n,m)}_\gamma}\ket{\oslash,\oslash}_{n,m}
        + \sqrt{p^{(n,m)}_\gamma}\ket{1,\oslash}_{n,m}\Big)\Bigg\}
    \end{split}
\end{equation}
where $p^{(n,m)}_\gamma$ is the probability of successfully generating a photon when creating the $m^{\rm th}$ qubit of the $n^{\rm th}$ photonic vertex of the state.
From this state is can be seen that inefficient excitation acts to modify the time-bin basis states of the photonic qubits without changing the entanglement between photonic qubits.
To see the impact of inefficient excitation on the photonic fusion process we need only consider two redundantly encoded vertices.
In this case the $f_X$ functions in Eq.~\eqref{eq:fusion_in} take the form $f_X=\sqrt{q}~\mathbb{I} + \sqrt{p}\mathcal{O}_{\rm in}^\dagger$ for $\mathcal{O}\in\{A,B,C,D\}$ such that for two redundantly encoded vertices
\begin{equation}
    \begin{split}
        \ket{\psi_{\rm in}} = &\frac{1}{2}\Big(f_A^{m-1}(\sqrt{q_1}\mathbb{I} + \sqrt{p_1}A^\dagger) + f_B(\sqrt{q_1}\mathbb{I} + \sqrt{p_1}B^\dagger)\Big)
        \otimes\Big(f_C^{m-1}(\sqrt{q_1}\mathbb{I} + \sqrt{p_1}C^\dagger) + f_D(\sqrt{q_1}\mathbb{I} + \sqrt{p_1}D^\dagger)\Big)
    \end{split}
\end{equation}
Performing the relevant transformation of the state reveals when inefficient excitation of the quantum emitters results in no photon being input into the fusion circuit there is no change in the wider state as expected.
However, when a single-photon is input into the fusion circuit without knowledge of which quantum emitter generated the photon entanglement can still successfully be generated.

Off-resonant excitation on the other hand results in the undesired emission of a photon co-polarised with the desired photon output but with a distinct energy.
Once again this results in a modification of the time-bin basis states, only now a multi-photon component is introduces rather than a vacuum component as can be seen in the form of the resulting resource state
\begin{equation}
\begin{split}
    \ket{\Phi_{\rm RE}}_N =& \mathbf{CZ}_{\uparrow,\mathcal{V}_N}^\prime \frac{1}{\sqrt{2}}\Big(\ket{\downarrow}\pm(-1)^N\ket{\uparrow}\Big)
    \Bigg(\prod_{n=1}^{N-1} \mathbf{CZ}_{\mathcal{V}_{n+1},\mathcal{V}_n}^\prime\Bigg)\\
    &\bigotimes_{n=1}^N\frac{1}{\sqrt{2}}\Bigg\{\bigotimes_{m=1}^{M_n}\Big(\sqrt{1-p^{(n,m)}_\uparrow}\ket{\oslash,\Tilde{H}}_{n,m}
    + \sqrt{p^{(n,m)}_\uparrow}\ket{H,\Tilde{H}}_{n,m}\Big)\\
    &+ \bigotimes_{m=1}^{M_n}\Big(\sqrt{1-p^{(n,m)}_\uparrow}\ket{\Tilde{H},\oslash}_{n,m}
    + \sqrt{p^{(n,m)}_\uparrow}\ket{\Tilde{H},H}_{n,m}\Big)\Bigg\}.
\end{split}
\end{equation}
Here we have moved from the photon number basis to the polarisation and energy (indicated by the presence of a tilde) of the emitted photons.
In this case, again focusing on entanglement generation only between two redundantly encoded vertices, the state input into the fusion circuit may be expressed as
\begin{equation}
    \begin{split}
        \ket{\psi_{\rm in}} = \frac{1}{2}\Big(&f_A^{m-1}(\sqrt{q_1}A_{\rm in}^\dagger + \sqrt{p_1}(A_{\rm in}^\dagger \bar{B}_{\rm in}^\dagger)
        + f_B^{m-1}(\sqrt{q_1}B_{\rm in}^\dagger + \sqrt{p_1}\bar{A}_{\rm in}^\dagger B_{\rm in}^\dagger)\Big)\\
        \otimes\Big(&f_C^{m-1}(\sqrt{q_2}C_{\rm in}^\dagger + \sqrt{p_2}C_{\rm in}^\dagger \bar{D}_{\rm in}^\dagger)
        + f_D^{m-1}(\sqrt{q_1}D_{\rm in}^\dagger + \sqrt{p_1}\bar{C}_{\rm in}^\dagger D_{\rm in}^\dagger)\Big)\ket{\oslash}.
    \end{split}
\end{equation}
where, for example, $f_A = \sqrt{q}A^\dagger + \sqrt{p}A^\dagger\bar{B}^\dagger$, the bar differentiates the frequencies of the photons generate by the two driven transitions, $p$ is the probability of exciting the off-resonant transition, and $q=1-p$.
Performing the unitary transformation applied by the fusion circuit reveals that the presence of additional photons from off-resonant excitation does not impact the fusion process directly assuming any additional photons are completely distinguishable from the desired photon states.
The main impact is to introduce uncertainty into the detector patterns used to herald successful entanglement generation.

\subsection{Photon loss}\label{app:photon_loss}

Perhaps the most significant error mechanism experienced by current optical systems is photon loss.
When we explicitly track the unique optical modes into which lost photons are scattered the time-bin basis states of the photonic qubits are once again modified only now accounting for probabilistic emission of photons into the desired optical mode and the loss modes.
The resulting entangled spin-photon resource state may be written as
\begin{equation}
\begin{split}
    \ket{\Phi_{\rm RE}}_N =& \mathbf{CZ}_{\uparrow,\mathcal{V}_N}\Bigg(\prod_{n=1}^{N-1}\mathbf{CZ}_{\mathcal{V}_{n+1},\mathcal{V}_n}\Bigg)
    \frac{1}{\sqrt{2}}\Big(\ket{\downarrow}\pm(-1)^N\ket{\uparrow}\Big)\bigotimes_{n=1}^{N}\frac{1}{\sqrt{2}}\Bigg\{\Bigg(\prod_{j=1}^{J_n-1}\mathcal{E}_{\mathcal{S}_{n,j+1},\mathcal{S}_{n,j}}\Bigg)\\
    &~~~~~~~~\bigotimes_{\mathcal{S}_{n,j}\in\mathcal{V}_n}\Bigg[\bigotimes_{m\in\mathcal{S}_{n,j}}\Bigg(\sqrt{q_\oslash^{(n,m)}}\ket{\oslash,\oslash}_{n,m}^{L}\ket{\oslash,1}_{n,m} + \sqrt{p_\oslash^{(n,m)}}\ket{\oslash,1}_{n,m}^{L}\ket{\oslash,\oslash}_{n,m}\Bigg)\\
    &~~~~~~~~~~~~~~~~~~~~~+\bigotimes_{m\in\mathcal{S}_{n,j}}\Bigg(\sqrt{\tilde{q}_\oslash^{(n,m)}}\ket{\oslash,\oslash}_{n,m}^{L'}\ket{1,\oslash}_{n,m} + \sqrt{\tilde{p}_\oslash^{(n,m)}}\ket{1,\oslash}_{n,m}^{L'}\ket{\oslash,\oslash}_{n,m}\Bigg)\Bigg]\Bigg\}\\
\end{split}
\end{equation}
As it is the basis states that are modified the entanglement between photonic qubits is unmodified with the entanglement operators now acting on the modified basis states.
Tracing over photons scattered into loss modes projects the system into a mixed state where entanglement within vertices redundantly encoded on GHZ states is completely destroyed.
The lack of entanglement after photon loss naturally prevents photonic fusion from generating entanglement between redundantly encoded vertices in the state(s).

\end{document}